%% file: main.tex
\documentclass[sigplan,twocolumn]{acmart}
\renewcommand\footnotetextcopyrightpermission[1]{}

\makeatletter
\AtBeginDocument{%
  \fancyhead[L]{\ACM@linecountL\@headfootfont}%
  \fancyhead[R]{\@headfootfont\ACM@linecountR}%
}
\makeatother

\input{macros}

\begin{document}

\title[Repeated Variational Circuit Optimization on GPUs]%
      {Making Repeated Variational Circuit Optimization Efficient on GPUs}

\author{Runzhe Zhang}
\authornote{Both authors contributed equally to this research.}
\affiliation{%
  \institution{Shanghai Jiao Tong University}
  \city{Shanghai}
  \country{China}
}
\email{zhang-rz@sjtu.edu.cn}

\author{Jiaqi Liu}
\authornotemark[1]
\affiliation{%
  \institution{Shanghai Jiao Tong University}
  \city{Shanghai}
  \country{China}
}
\email{ofnoname@sjtu.edu.cn}

\author{Anbang Wu}
\affiliation{%
  \institution{Shanghai Jiao Tong University}
  \city{Shanghai}
  \country{China}
}
\email{anbang@cs.sjtu.edu.cn}

\author{Pu Pang}
\affiliation{%
  \institution{Shanghai Jiao Tong University}
  \city{Shanghai}
  \country{China}
}
\email{pangpu@cs.sjtu.edu.cn}

\author{Shixuan Sun}
\affiliation{%
  \institution{Shanghai Jiao Tong University}
  \city{Shanghai}
  \country{China}
}
\email{sunshixuan@sjtu.edu.cn}

\author{Yu Feng}
\affiliation{%
  \institution{Shanghai Jiao Tong University}
  \city{Shanghai}
  \country{China}
}
\email{y-feng@sjtu.edu.cn}

\author{Quan Chen}
\affiliation{%
  \institution{Shanghai Jiao Tong University}
  \city{Shanghai}
  \country{China}
}
\email{chen-quan@cs.sjtu.edu.cn}

\author{Jingwen Leng}
\affiliation{%
  \institution{Shanghai Jiao Tong University}
  \city{Shanghai}
  \country{China}
}
\email{leng-jw@cs.sjtu.edu.cn}

\author{Minyi Guo}
\affiliation{%
  \institution{Shanghai Jiao Tong University}
  \city{Shanghai}
  \country{China}
}
\email{guo-my@cs.sjtu.edu.cn}



\input{sections/00-abstract}



\maketitle

\input{sections/01-introduction}
\input{sections/02-background}

\input{sections/03-characterization}

\input{sections/05-qark}
\input{sections/08-implementation}
\input{sections/09-evaluation}

\input{sections/10-related-work}

\input{sections/12-conclusion}

\section*{Acknowledgement}
ChatGPT was utilized to polish this work, including text, figures and tables.

\bibliographystyle{ACM-Reference-Format}
\bibliography{references}

\end{document}

%% file: macros.tex
\usepackage{amsmath}
\usepackage{mathtools}
\usepackage{graphicx}
\usepackage{multirow}
\usepackage{xspace}
\usepackage{xcolor}
\usepackage{colortbl}
\usepackage{tabularx}
\usepackage{siunitx}
\usepackage{algorithm}
\usepackage{subcaption} 
\usepackage{algpseudocode}
\usepackage{comment}
\usepackage[capitalise,nameinlink,noabbrev]{cleveref}
\usepackage[most]{tcolorbox}

\newcommand{\system}{\textsc{QARK}\xspace}
\newcommand{\baseline}{PL-GPU\xspace}

\newcommand{\eg}{e.g.,\xspace}

\newcommand{\fillin}[1]{\textcolor{red!70!black}{\textbf{[#1]}}}

\newcommand{\placeholderbox}[3]{%
  \fbox{%
    \begin{minipage}[c][#2][c]{#1}
      \centering\small #3
    \end{minipage}}}

\newcommand{\ket}[1]{\lvert #1\rangle}
\newcommand{\bra}[1]{\langle #1\rvert}

\newcommand{\loss}{\mathcal{L}}

\crefname{section}{Section}{Sections}
\Crefname{section}{Section}{Sections}
\crefname{figure}{Figure}{Figures}
\Crefname{figure}{Figure}{Figures}
\crefname{table}{Table}{Tables}
\Crefname{table}{Table}{Tables}
\crefname{equation}{Equation}{Equations}
\Crefname{equation}{Equation}{Equations}
\crefname{algorithm}{Algorithm}{Algorithms}
\Crefname{algorithm}{Algorithm}{Algorithms}

\newtcolorbox{takeawaybox}{
  colback=yellow!10,     
  colframe=black,        
  boxrule=0.4pt,
  arc=0pt,
  left=4pt,
  right=4pt,
  top=3pt,
  bottom=3pt,
  before skip=6pt,
  after skip=6pt,
  width=\linewidth,
  breakable
}

\newcommand{\pp}[1]{{\textcolor{magenta}{ #1}}}

%% file: sections/00-abstract.tex
\begin{abstract}
Repeated numerical optimization is a defining workload for classical simulation of parameterized quantum circuits. Although successive training steps change gate parameters, they usually preserve circuit connectivity and reuse a small set of gate-layer patterns. We use this regularity to accelerate gradient evaluation on GPUs. Our framework combines execution primitives for common circuit layers with adjoint differentiation, coordinating state evolution and derivative computation to reduce launch overhead, intermediate storage, and state-vector traffic. Because effective schedules depend on circuit size, gate structure, and device resources, our framework retains several execution choices instead of committing to a single schedule. Structural and hardware constraints narrow these choices, after which measurements on representative layers guide plan selection. The selected plans are reused while circuit structure remains fixed, spreading the selection cost over repeated training iterations. This approach targets the computation required by optimization rather than forward simulation alone, while retaining the numerical semantics of state-vector differentiation. Experiments across multiple circuit families and GPU platforms compare our framework with the native GPU adjoint implementation in PennyLane Lightning. The results demonstrate substantially lower gradient evaluation times over the evaluated workloads, together with the numerical accuracy checks reported in the evaluation.
\end{abstract}

%% file: sections/01-introduction.tex
\section{Introduction}
\label{sec:introduction}

A variational quantum experiment repeatedly solves the same computational problem with changing numerical inputs: evaluate a circuit, obtain derivatives of an objective, and update its parameters. Classical simulation makes this process accessible when quantum processors are unavailable or too noisy for controlled experiments. It supports the development of quantum learning models and variational algorithms before deployment on hardware~\cite{perez2020data,Peruzzo2014A,qaoa}. For these applications, the relevant performance target is the cost of repeated optimization steps, including their derivatives.

This workload has useful regularity. The gate connectivity and operation sequence normally remain unchanged throughout training, while parameter values evolve. Many circuit families also repeat a small collection of rotation and entangling layers. These layers need not execute simultaneously on quantum hardware to be useful computational units in a simulator. Their known structure allows a GPU implementation to organize state-vector accesses and combine work across gates, reducing repeated launches and transfers to device memory.

Differentiation makes exploiting this regularity more demanding. Evaluating each parameter independently can require many circuit executions, while retaining all intermediate states for reverse-mode differentiation consumes substantial memory. For reversible unitary circuits, adjoint differentiation reconstructs intermediate states during a reverse traversal and avoids storing the complete execution history~\cite{schuld2019parameter,jones2020adjoint,luo2020yao}. Its favorable traversal count does not make GPU execution automatically efficient: derivative contractions, state reconstruction, and temporary storage still determine how much parallelism and data reuse an implementation can realize.

Our framework organizes simulation around the repeated training workload. It provides GPU execution choices for both state evolution and adjoint gradient evaluation, including operations on common gate layers. Independent derivative work can be parallelized, and partial gradient contributions can be accumulated without retaining every intermediate result. The resulting schedules offer different balances among launch frequency, working memory, and state-vector traffic. This makes the implementation responsive to circuit structure while preserving the numerical computation required by training.

Choosing among these schedules is an empirical problem. A configuration that benefits a small circuit may expose insufficient parallelism at another size, and a schedule with good locality may demand too many resources on a different GPU. A purely analytical performance prediction is difficult across these cases. Our framework uses structural restrictions and GPU resource constraints to eliminate unsuitable choices, then measures promising alternatives on the target device. Repeated gate layers provide compact representatives for this process, and fixed circuit structure allows the selected plans to be reused across optimization iterations. Thus, measurements guide execution without requiring a full search at every training step.

GPU adjoint simulation is already supported by existing software, including PennyLane Lightning~\cite{asadi2024,pennylane-support-adjoint}. Our focus is the additional benefit of coordinating layer execution, derivative computation, and plan selection for repeated optimization. We evaluate our framework using gradient evaluation time across several circuit families, circuit sizes, and GPUs, with the native GPU adjoint implementation in PennyLane Lightning as the baseline. The reported experiments show substantial reductions in this cost. The subsequent sections characterize the workloads, describe the execution choices and selection procedure, and examine performance and numerical accuracy.

%% file: sections/02-background.tex
\section{Background}
\label{sec:bg}

This section introduces necessary concepts for parameterized quantum circuit (PQC) training.

\subsection{Quantum Computing Basics}

A quantum bit, i.e., a qubit, is a unit vector in
the two-dimensional complex Hilbert space,
\begin{equation}
  \ket{\phi}=\alpha\ket{0}+\beta\ket{1},
  \qquad |\alpha|^2+|\beta|^2=1,
  \label{eq:single-qubit-state}
\end{equation}
where the complex amplitudes $\alpha$ and $\beta$ determine the probabilities
of observing the qubit in states $\ket{0}$ and $\ket{1}$, respectively.
More generally, the state of an
$n$-qubit system is represented by a state vector,
\begin{equation}
  \ket{\phi}=\sum_{x=0}^{2^n-1} a_x\ket{x},
  \qquad \sum_{x=0}^{2^n-1}|a_x|^2=1,
  \label{eq:n-qubit-state}
\end{equation}
whose $N=2^n$ amplitudes each associate a complex value with a computational
basis state $\ket{x}$.
Thus, an exact state-vector representation on
a classical machine requires memory that grows
exponentially with the number of qubits.

A quantum gate acting on a $k$-qubit state is represented
by a $2^k\times 2^k$ unitary matrix $U$. Commonly used
quantum gates include:
\begingroup
\setlength{\arraycolsep}{2pt}
\begin{align}
  \mathrm{CX}
  =
  \begin{pmatrix}
    \mathrm{I} & \textbf{0} \\
    \textbf{0} & \mathrm{X}
  \end{pmatrix}, \, 
  \mathrm{X}
  =
  \begin{pmatrix}
    0 & 1 \\
    1 & 0
  \end{pmatrix}, \quad 
  R_y(\theta) =
  \begin{pmatrix}
    C_\theta & \ -S_\theta \\
    S_\theta & \ C_\theta
  \end{pmatrix}, \label{eq:common-quantum-gates}
  \\\nonumber
  R_x(\theta)
  =
  \begin{pmatrix}
    C_\theta & -iS_\theta \\
    -iS_\theta & C_\theta
  \end{pmatrix}, \quad
  R_z(\theta) =
  \begin{pmatrix}
    e^{-i\theta} & 0 \\
    0 & e^{i\theta}
  \end{pmatrix},
\end{align}
\endgroup
where $C_\theta = \cos(\theta/2)$ and $S_\theta = \sin(\theta/2)$. Other quantum gates
can be decomposed into these gates,
e.g., the Hadamard gate $H=R_y(\pi)R_z(\pi)$, 
the phase gates $S=e^{-i\frac{\pi}{2}}R_z(\frac{pi}{2})$, 
and $T=e^{-i\frac{\pi}{8}}R_z(\frac{pi}{4})$.

A quantum circuit consists of a sequence of such gates;
its final state is obtained by applying the
corresponding gates to the initial state in circuit order.
A circuit is called a parameterized quantum circuit (PQC) if some of its gates
contain undetermined parameters.

\subsection{Quantum Circuit Simulation}
Because access to real quantum hardware is limited and current devices
remain noise-prone, classical circuit simulators are widely used for quantum
algorithm development, debugging, and validation.
Existing work accelerates quantum circuit simulation (QCS) along  
dimensions: optimizing individual gate kernels, reducing kernel
launches, improving memory access and exploiting hardware parallelism~\cite{qgpu2022,atlas2024,cast2025}.

\begin{figure}[t]
  \centering
    \includegraphics[width=0.95\linewidth]{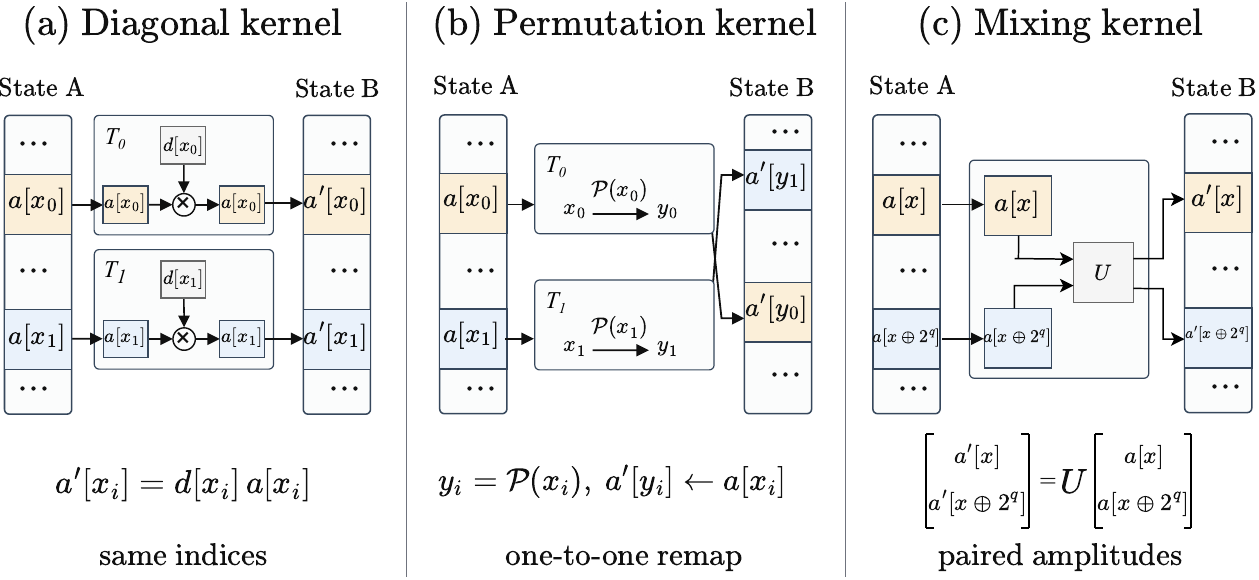}
  \caption{Optimized GPU kernels for various quantum gates.}
  \label{fig:bgkernel}
\end{figure}

\textbf{Gate kernel optimization.}
Operators can be classified by how they transform amplitudes in the
computational basis, enabling specialized gate kernels.
As illustrated in Figure~\ref{fig:bgkernel},
\textsc{diagonal operators} update each amplitude independently and
can be applied in a single traversal of the state vector.
\textsc{Permutation or index swapping operators}, such as CX and SWAP gates, move
amplitudes between basis indices. 
\textsc{Mixing operators}, such as $R_x$, $R_y$, and Hadamard gates,
form linear combinations of amplitudes.

\textbf{Kernel launch optimization.}
\textsc{Gate fusion} is a representative technique for reducing
kernel launches. It combines consecutive gates into a single
execution unit. In its most general form, a series of gates $U_i,\ldots,U_j$
are replaced by a single composite operator $V=U_j\cdots U_i$.
However, larger fused regions can increase the composite operator
size and thus register or memory pressure for related computation.


\textbf{Memory access design.}
When the state vector exceeds the on-chip storage capacity, it can
be processed in tiles. Two commonly used layouts differ in how
qubit bits are mapped to tile addresses.
The \textsc{low-order layout} prioritizes low-order qubit bits,
favoring contiguous reads.
The \textsc{high-order layout} instead prioritizes high-order
qubit bits, facilitating the execution of gates acting on
high-order qubits.

\textbf{GPU parallelism.} A typical state-vector simulator stores the amplitudes $a_x$ of an $n$-qubit quantum state in a contiguous array, where the bits of index
$x$ identify the corresponding computational basis state $\ket{x}$.
Applying a $k$-qubit gate partitions this array into $2^{n-k}$ independent
batch of $2^k$ amplitudes. Within each batch, the indices differ only
at the target-qubit bit positions, and each batch is updated by the
same $2^k \times 2^k$ matrix-vector multiplication.
On a GPU, these batches are distributed across many threads, which
cooperatively load, transform, and write the amplitudes, exposing
abundant data parallelism.

Although these strategies are commonly used in general-purpose
QCS, adapting them to PQC training simulation is nontrivial, due to the additional gradient computation stage.



\subsection{PQC Training}

A PQC is usually of the form:
\begin{equation}
  U(\boldsymbol{\theta})
  =
  U_P(\theta_P)\cdots U_2(\theta_2)U_1(\theta_1),
  \label{eq:pqc}
\end{equation}
which transforms an initial state \(\ket{\phi_0}\), typically
\(\ket{0}^{\otimes n}\), into the final state
\(\ket{\phi_P}=U(\boldsymbol{\theta})\ket{\phi_0}\).
A common PQC loss function is the expectation value of a Hermitian observable \(H\):
\begin{equation}
  \loss(\boldsymbol{\theta})
  =
  \bra{\phi_P}H\ket{\phi_P}
  = \bra{\phi_0} U^\dagger(\boldsymbol{\theta})\cdot H
  \cdot U(\boldsymbol{\theta})\ket{\phi_0}.
  \label{eq:objective}
\end{equation}

Training the PQC repeatedly evaluates \(\loss\) and its gradient
\(\nabla_{\boldsymbol{\theta}}\loss\)~\cite{Peruzzo2014A}. 
There are multiple ways to compute \(\nabla_{\boldsymbol{\theta}}\loss\); 
we describe the three most representative ones here.

\textbf{Parameter-shift rule (PSR) \cite{schuld2019parameter}.}
PSR evaluates the derivative as
    $\frac{\partial\loss}{\partial\theta_i}
  =
  c\!\left[
    \loss(\theta_i+s)-\loss(\theta_i-s)
  \right],$
where \(s\) and \(c\) are determined by the gate parameterized by \(\theta_i\) \cite{schuld2019parameter}. 
$\loss(\theta_i+s)$ (or $\loss(\theta_i-s)$) denotes evaluating the
loss function (and circuit) with $\theta_i$ replaced by $\theta_i+s$ (or $\theta_i-s$).
Therefore, PSR requires two circuit evaluations to compute the derivative of a single parameter.
For a PQC with \(P\) parameters, computing the full gradient requires \(2P\) circuit
evaluations. When running PSR on a state-vector simulator,
each circuit evaluation stores one \(N=2^n\)-amplitude state to compute the loss, consuming \(O(2^n)\) memory.

\textbf{Back propagation (BP) \cite{Rumelhart1986Learning}.}
BP regards quantum gates as matrices and states as vectors, viewing
the PQC as a differentiable program made of complex 
matrix-vector operations.  BP then applies
the chain rule to propagate derivatives from the loss back to the parameters:
\begin{center}
    $
  \frac{\partial \loss}{\partial \theta_i} = 
  \sum_{j=1}^P \sum_{r=1}^{2^n} \sum_{c=1}^{2^n}
  \frac{\partial \loss}{\partial u_{j,r,c}}  
  \frac{\partial u_{j,r,c}}{\partial \theta_i} 
  + \frac{\partial \loss}{\partial u_{j,r,c}^\dagger} 
  \frac{\partial u_{j,r,c}^\dagger}{\partial \theta_i},
$
\end{center}
where $u_{j,r,c}$ is the element of $U_j$ at the $r$-th row and $c$-th column.
The intermediate derivatives $\{\partial \loss / \partial u_{j,r,c}\}$
are computed and stored via
the chain rule along the computation graph.
By reusing these stored intermediate derivatives, BP avoids launching
extra circuit evaluations for every parameter.
Typically, BP requires only one forward traversal and
one backward traversal of the PQC to compute the whole gradient.
However, BP consumes $O(P4^n)$ memory to
store these intermediate derivatives, for an \(n\)-qubit PQC.

\begin{figure}[t]
  \centering
    \includegraphics[width=0.48\textwidth]{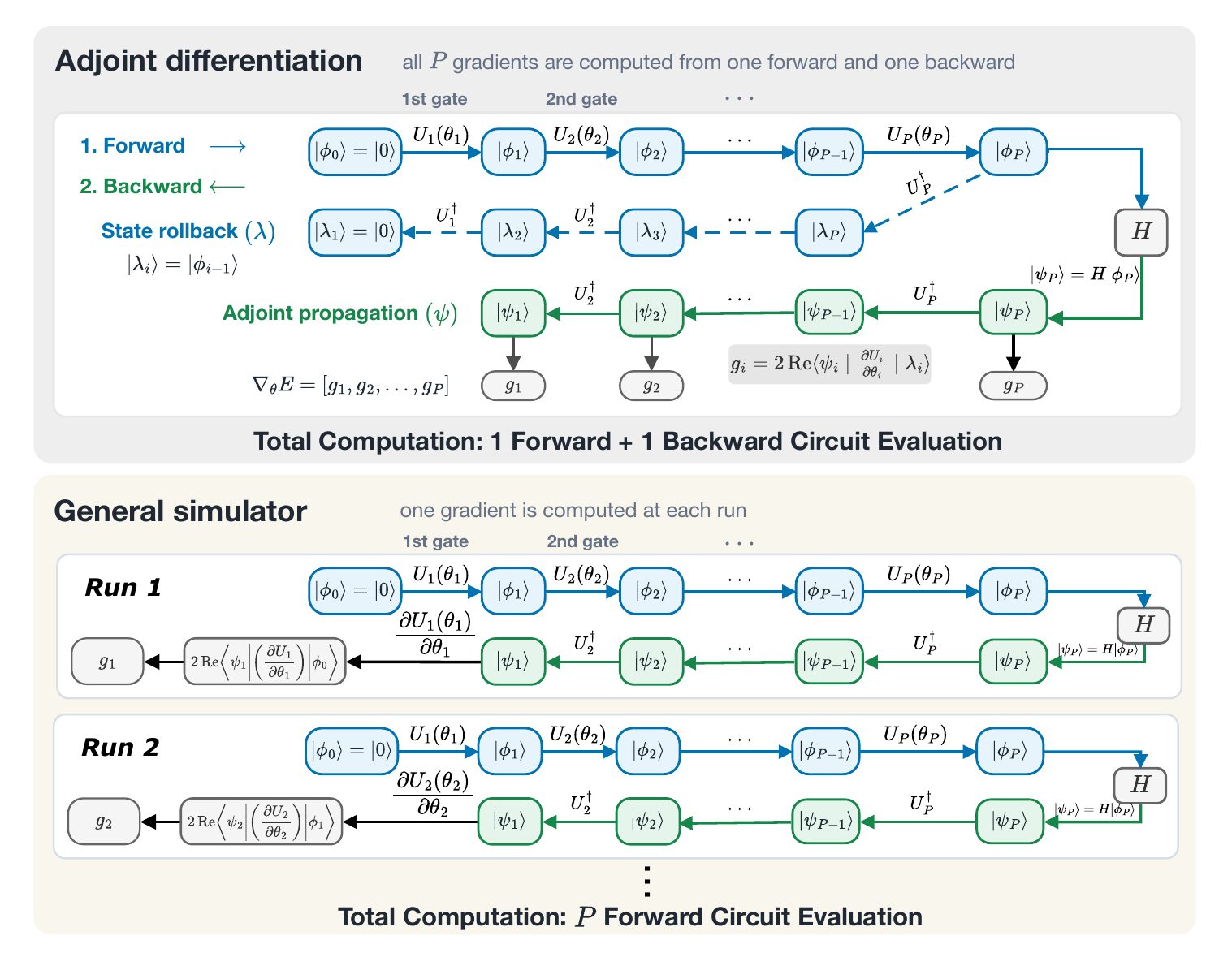}
  \caption{The Adiff gradient computation pipeline. General simulators can be adapted to run the Adiff pipeline, but at substantial overhead.}
  \label{fig:adiffpipe}
\end{figure}

\textbf{Adjoint differentiation (Adiff) \cite{jones2020adjoint,luo2020yao}.}
Adiff instead stores only two statevectors, 
$\ket{\psi_i}$ and $\ket{\lambda_i}$, as the intermediate
quantities for computing derivatives:
\begin{align}\label{eq:gradient}
  \frac{\partial\loss}{\partial\theta_i}
  &= \bra{\phi_0} \cdots \frac{\partial U_i^\dagger}{\partial \theta_i}
  \cdots U_P^\dagger H U_P \cdots \ket{\phi_0}\  + \\[-2pt] \nonumber
  & \bra{\phi_0} 
  \cdots U_P^\dagger H U_P \cdots
  \frac{\partial U_i}{\partial \theta_i} \cdots \ket{\phi_0}
  = 2\mathrm{Re} \bra{\psi_i} \frac{\partial U_i}{\partial \theta_i} \ket{\lambda_{i}},
\end{align}
where $\ket{\psi_{P}} = H \ket{\phi_P}$,
$\ket{\lambda_{P}} = U_P^\dagger \ket{\phi_P}$,
$\ket{\psi_{i}} = U^\dagger_{i+1}\ket{\psi_{i+1}}$,
$\ket{\lambda_{i}}=U_{i}^\dagger\ket{\lambda_{i+1}}$. We can compute
$\ket{\lambda_P}$ and $\ket{\psi_P}$ by a forward traversal of the circuit,
and then compute $\ket{\lambda_i}$ and $\ket{\psi_i}$ by a backward traversal.
Consequently, Adiff computes the loss and the complete gradient with only two
traversals of the PQC, rather than
\(O(P)\) independent circuit evaluations. 
Adiff consumes \(O(2^n)\) memory to keep
two intermediate statevectors $\ket{\psi_i}$ and $\ket{\lambda_i}$
\cite{jones2020adjoint,luo2020yao}.
Figure \ref{fig:adiffpipe} summarizes the Adiff algorithm. 

In this work, we focus on optimizing Adiff,
as it is currently the standard approach for PQC training,
due to its superiority over the other two methods \cite{jones2020adjoint,luo2020yao}.
We also focus on the statevector-based simulation of Adiff,
since Adiff is naturally developed in this form.
Explicit tracking of state vectors
enables us to quickly debug Adiff's behavior.

%% file: sections/03-characterization.tex
%

\begin{figure*}[t]
  \centering
  \includegraphics[width=0.95\textwidth]{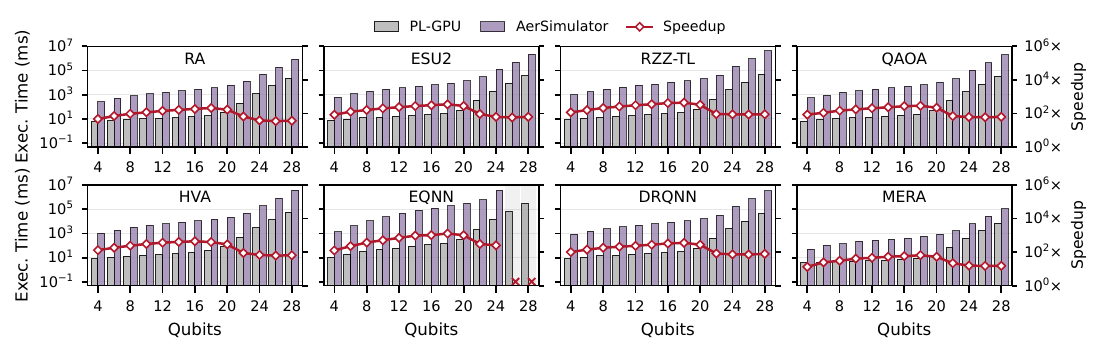}
  \caption{Comparison of \baseline{} (PennyLane-Lightning-GPU simulator) with Qiskit AerSimulator on RTX 6000 Ada in terms of Adiff computation execution time. The speedup is computed by ‘Execution time of AerSimulator / Execution time of \baseline{}’. As AerSimulator fails to produce results within 12 hours for EQNN circuits with more than 24 qubits, its results are omitted here.}
  \label{fig:qiskit-vs-PL}
\end{figure*}

\section{Motivation}\label{sec:moti}

In this section, we evaluate the performance of existing quantum circuit simulators for Adiff computation, seeking to answer two questions: (1) Are existing simulators efficient enough for PQC training simulation? (2) If not, what are the root causes of the inefficiency?

\subsection{Investigation Setup}

We conduct the investigation on two servers, one with an NVIDIA RTX 6000 Ada GPU and the other with an NVIDIA RTX 4090 GPU.
The investigation uses eight PQCs with 4, 6, 8, ..., 26, and 28 qubits, which are commonly used in prior studies of quantum circuit simulation~\cite{lykov2021performance,efthymiou2022qibo,suzuki2021qulacs,efthymiou2022quantum,tensorcircuit2023,zhang2025bmqsim}, including three Hardware-Efficient Ansatz (HEA) circuits from Qiskit Circuit Library~\cite{qiskit-circuit-library}---RealAmplitudes (RA), EfficientSU2 (ESU2) and RZZ-entangled TwoLocal (RZZ-TL), Quantum Approximate Optimization Algorithm (QAOA) on MAXCUT problems~\cite{qaoa},  XXZ Hamiltonian Variational Ansatz (HVA)~\cite{PRXQuantum.1.020319}, Equivariant Quantum Neural Network (EQNN)~\cite{PRXQuantum.4.010328,Schatzki2024}, Data Re-uploading Quantum Neural Network (DRQNN)~\cite{perez2020data}, and Multi-scale Entanglement Renormalization Ansatz (MERA)~\cite{grant2018hierarchical}.
More details about the experimental platforms and circuits are provided in Section~\ref{sec:evaluation-methodology}.

In terms of the quantum circuit simulators, two simulators based on NVIDIA cuQuantum~\cite{cuquantum2023} are used: the general-purpose quantum circuit simulator AerSimulator of Qiskit Aer 0.17.1~\cite{qiskit-aer} and \texttt{lightning.gpu}~0.45.0 of PennyLane~\cite{pennylane-lightning-docs} (denoted as ``PL-GPU'') which provides dedicated support for Adiff~\cite{asadi2024, pennylane-support-adjoint}.

\subsection{Inefficiencies of Existing Simulators}\label{sec:PL-vs-Qiskit}

General-purpose QCS simulators lack native support for backward gradient computation. Although they can execute Adiff by recasting the backward computation as a forward pass, this adaptation incurs substantial overhead. As illustrated in Figure~\ref{fig:adiffpipe}, this overhead arises primarily from their inability to reuse intermediate state vectors to reduce the number of circuit evaluations.


Experiments in Figure~\ref{fig:qiskit-vs-PL} further justify this observed inefficiency. Figure~\ref{fig:qiskit-vs-PL} compares the performance of AerSimulator and PL-GPU on an NVIDIA RTX 6000 Ada GPU, regarding gradient computation. For each PQC circuit, we execute Adiff 1000 times and report the mean end-to-end execution time, measured from the first GPU kernel launch until the final result is available on the CPU. The red curve, plotted against the secondary y-axis, shows the speedup of PL-GPU over AerSimulator. 

As shown in Figure~\ref{fig:qiskit-vs-PL}, AerSimulator incurs prohibitively long execution times for Adiff, averaging 264.9s.
For EQNNs with more than 24 qubits, the execution time even exceeds 12 hours.
With the dedicated Adiff support, PL-GPU achieves a geometric-mean speedup of 134.8$\times$.
Similar speedups are also observed on the RTX 4090 GPU.
The speedup is mainly attributed to general QCS simulator cannot
reduce circuit evaluations by reusing intermediate state vectors, as explained in Figure \ref{fig:adiffpipe}.

However, despite the significant speedup in Adiff, simulating the overall PQC training with PL-GPU remains time-consuming. 
This is because training a PQC typically requires thousands to tens of thousands of iterations~\cite{wang2022qoc, meyer2023quantum,zhang2026aqer}, each of which involves an Adiff, resulting in considerable overall simulation time on GPUs if the Adiff execution time is not sufficiently low.
This issue becomes more pronounced as the number of qubits increases, since larger circuits generally require more training iterations. 

For example, training a 22-qubit EQNN requires more than 10000 iterations. 
Although PL-GPU reduces the execution time from 628s (AerSimulator) to 2s, the overall simulated training time still reaches at least 2s $\times$ 10000 = 20000s (5.56 hours).
Even for a 10-qubit EQNN, the simulation with PL-GPU still takes 7.7 minutes.

\begin{takeawaybox}
\textbf{Takeaway.}
The prohibitively long Adiff execution time on the general-purpose simulator highlights the need for dedicated support. However, the existing simulator with such support still does not provide sufficient performance for efficient PQC training simulation, motivating further optimization. 
\end{takeawaybox}




\subsection{Root Causes of the Inefficiency}\label{sec:root_causes}

To understand \baseline{}'s inefficiency, we first break down overall execution time to identify where most of the time is spent; we then examine GPU performance metrics to characterize how PL-GPU utilizes the GPU in Adiff computation.

{\bf Execution time breakdown.} 
When using \baseline{} for an Adiff, the execution time can be divided into two parts: (1) the forward traversal of the circuit (denoted as ``Forward''), and (2) the reverse traversal of the circuit (``Backward'').

For the eight PQCs, we measure the time of the two parts at different qubit counts and compute their proportions of the overall execution time. For a given qubit count, these proportions are close across different circuits and the two GPUs (RTX 6000 Ada and RTX 4090).

\begin{figure}[t]
  \centering
    \includegraphics[width=0.35\textwidth]{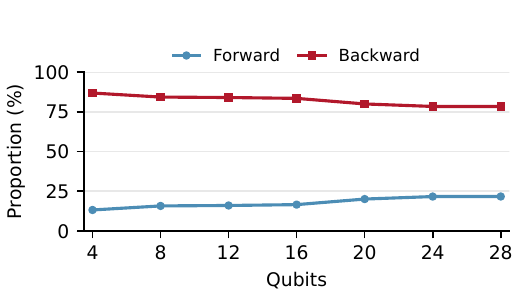}
  \caption{The breakdown of Adiff execution time as the number of qubits varies when using \baseline{}.}
  \label{fig:baseline_breakdown}
\end{figure}

Figure~\ref{fig:baseline_breakdown} shows the average proportions across the eight circuits on the RTX 6000 Ada GPU for given qubit count.
As shown, {\bf Backward dominates the Adiff}, contributing 78.3\%$\sim$86.8\%.
It is as expected: the backward gradient computation needs to evolve two state vectors (see Equation \ref{eq:gradient}) while the forward pass only concerns one, let alone the backward pass also computes inner products of state vectors.

{\bf GPU utilization.} Using NVIDIA Nsight Systems~\cite{nsys} and NVIDIA Nsight Compute~\cite{ncu}, we profile PL-GPU during Adiff computation and report three metrics: (1) the ratio of the GPU active time, defined as the GPU non-idle time divided by the overall execution time~\cite{tan2019efficientnet, kwon2020nimble}, which reflects kernel-launch and synchronization overheads; (2) the GPU L2 cache hit rate; and (3) the GPU memory bandwidth. The latter two metrics characterize the memory-access behavior.

\begin{figure}[t]
  \centering
    \includegraphics[width=0.48\textwidth]{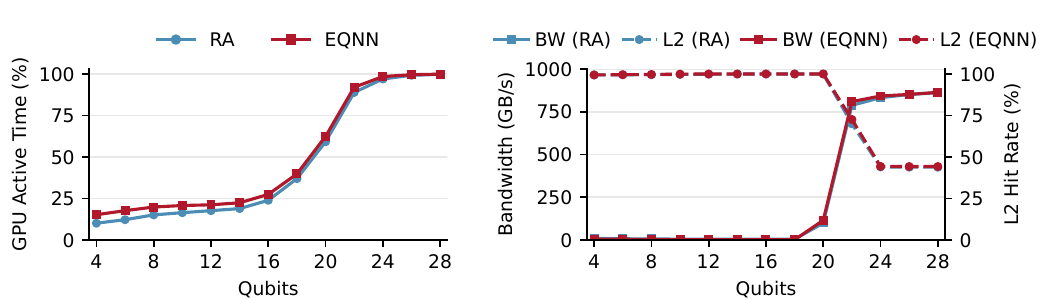}
  \caption{The left part of the figure shows the GPU active time ratio, while the right part shows the L2 cache hit rate and GPU memory bandwidth (BW).}
  \label{fig:baseline_GPU_metrics}
\end{figure}

Figure~\ref{fig:baseline_GPU_metrics} shows the three metrics for RA and EQNN across different qubit counts on the 6000 Ada GPU.
As shown, both the values and trends of the three metrics are very close between the two circuits. The other circuits also exhibit close values and trends for these metrics.

This is because PL-GPU {\bf uses a fixed GPU kernel schedule for any circuit}~\cite{pl-gpu-source-code}.
However, as shown in Figure~\ref{fig:benchmark-circuit}, RA and EQNN differ considerably in circuit structure and gate type. 
A fixed execution scheme is therefore unlikely to perform well for both; as shown in our experiments (Figure~\ref{fig:main-circuits} and~\ref{fig:platform-circuits}), their optimization potentials differ markedly.

\begin{takeawaybox}
\textbf{Takeaway.}
Existing simulator spends most of the Adiff execution time on the reverse traversal, while its fixed GPU kernel schedules leave circuit-specific optimization opportunities unexplored.
\end{takeawaybox}

%% file: sections/05-qark.tex
\section{Adiff-aware Tunable GPU Kernel Design}\label{sec:kdesign}

Adiff introduces
gradient computation that conventional QCS optimizations
do not directly address. We therefore develop derivative-aware kernels
and expose their execution choices for tuning across circuits and GPUs,
whose performance characteristics vary as discussed in \cref{sec:moti}.

\subsection{Derivative-aware Kernel Design}
\label{sec:comp}

Gate fusion is the core technique for reducing kernel launches. 
In general QCS, fusion only considers gate forwarding, while
Adiff must additionally recover every parameter's gradient 
within the fused region, which is shown to dominate the overall
Adiff simulation (see \cref{sec:moti}).
Our kernel designs account for efficient gradient computation,
exploiting operator structure, parallel derivative computation, and
derivative-aware memory access design.

\textbf{Adiff-compatible operator kernels.}
Gate derivatives retain exploitable matrix structures, 
preserving the diagonal
structure of $R_z$ and the amplitude-mixing patterns of $R_x$ and $R_y$.
Under the rotation convention in \cref{sec:bg}, we have 
$$
  \frac{\partial R_a(f(\theta))}{\partial\theta}
  = \frac{f'(\theta)}{2}R_a(f(\theta)+\pi),
  \qquad a\in\{x,y,z\}.
$$
Although derivative operators need not be unitary, their structure
allows us to reuse of the specialized diagonal and mixing kernels described
in \cref{sec:bg}, with the appropriate chain-rule factors.

\textbf{Derivative-aware fusion.} Consider the PQC defined in Equation \ref{eq:pqc}, and define the fused operator: 
$$V_{i:j}=U_j(\theta_j)\cdots U_i(\theta_i).$$ 
Considering computing $k$ derivatives for $\{\theta_{j-l}\}_{l=0}^{k-1}$ in parallel. Let $W=V_{j-k+1:j}$ and $E_l=\partial W/\partial\theta_l$.
Initialize a base buffer $s_0=\bra{\psi_j}$, one copy $s_l=s_0$
per derivative (aka derivative states), and $r_0=\ket{\lambda_{j+1}}$.
We then compute derivative states and update base buffer in parallel:
\begin{align}\label{eq:derive}
s_l &\leftarrow s_l E_l,\\ \nonumber
r_0 &\leftarrow W^\dagger r_0=\ket{\lambda_{j-k+1}},\\
s_0 &\leftarrow s_0 W=\bra{\psi_{j-k}} \nonumber
\end{align}
Afterwards, we compute
\(\frac{\partial\loss}{\partial\theta_l}=2\mathrm{Re}(s_l r_0).\)
The updated base buffers provide the inputs for following derivative computation.
This execution pattern uses two base-state buffers and up to $k$
derivative buffers.

The tunable choices are the gates assigned to each fusion group 
and the number of derivatives computed in parallel.

For shared parameters, we maintain one gradient entry per unique
parameter and accumulate all gate-level contributions into that entry. 

\textsc{Fusion tradeoffs.}
Parallel derivative computation trades storage for reduced execution
overhead. 
Let $T_{\mathrm{gate}}$ denote the cost of one gate application and
$T_{\mathrm{aux}}$ the overhead of buffer-copy and computing \(\frac{\partial\loss}{\partial\theta_l}=2\mathrm{Re}(s_l r_0)\).
Under ideal parallel execution, ignoring memory traffic and resource
contention, computing $s_l E_l$ costs approximately
$kT_{\mathrm{gate}}$, giving the full gradient computation time 
in the fusion group as
\(
  T_f(k)
  \approx \frac{j-i}{k}
    \left(kT_{\mathrm{gate}}+T_{\mathrm{aux}}\right).
\)
Increasing $k$ amortizes $T_{\mathrm{aux}}$, which is comparable
to $T_{\mathrm{gate}}$. But this requires more storage.

\textbf{Adiff-compatible tiling.}
Tiling limits the working set of exponentially growing state vectors.
However, for the differentiation example in Equation \ref{eq:derive},
general QCS tiling strategy would allocate every
$s_l$ state in global memory: it loads a tile of $s_l$, applies operators,
and writes the result back to $s_l$ before evaluating the inner product $s_l r_0$.
With $k$ parallel derivatives, this requires $k$ additional full-state
buffers, which is quite expensive.

We avoid this storage overhead by computing the inner product $s_l r_0$
even if $s_l$ state is partially known. Let
$\{f_\mathfrak{t}\}_{\mathfrak{t}=1}^{m}$ partition the state into disjoint tiles of
$2^n/m$ amplitudes, and define base-state tiles
\(
  v_\mathfrak{t}=f_\mathfrak{t}(\bra{\psi_j}),\quad
  u_\mathfrak{t}=f_\mathfrak{t}(\ket{\lambda_{j-k+1}}).
\)

If $E_l$ preserve each tile, i.e., $f_\mathfrak{t}(s_l E_l) = f_\mathfrak{t}(s_l) E_l$, then
\begin{equation}
  \frac{\partial\loss}{\partial\theta_l}
  = 2\mathrm{Re}\sum_{\mathfrak{t}=1}^{m}f_\mathfrak{t}(s_l E_l)f_\mathfrak{t}(\ket{\lambda_{j-k+1}})
  =2\mathrm{Re}\sum_{\mathfrak{t}=1}^{m}v_\mathfrak{t}E_{l}u_\mathfrak{t}.
  \label{eq:shard}
\end{equation}
Each kernel loads the required base-state tiles into registers or
shared memory, computes the derivative tiles, and accumulates their
inner products. Only the updated base states and partial gradient
sums need to leave the kernel; full derivative states need not be
stored in global memory.

We further expose tile size, tile mapping, and
the number of tiles processed concurrently per SM as tuning choices.


\input{figures/qark_kernel_options}

\subsection{Overall PQC Kernel Tuning Space}\label{sec:ova}

As shown in Table \ref{tab:qark-kernel-options},
beyond Adiff-aware designs in Sections~\ref{sec:comp}, 
we also expose kernel options studied in existing works, as
long as they affect PQC simulation.
We argue that these options still require Adiff-specific
tuning, as gradient computation changes performance tradeoffs.
For example, according to \cref{sec:comp},
computing $k$ derivatives in parallel requires $k+2$ state-tiles, 
so limited SM shared memory forces smaller tiles as $k$
increases, indicating Adiff imposes more restrictive 
storage designs than general QCS. 

The proposed Adiff-aware configuration space covers major optimization dimensions for efficient GPU simulation:

\textbf{Gate kernel and launch optimizations.} See \cref{sec:comp}.

\textbf{GPU parallelism options.}
We parameterize three controls: \emph{amplitudes per thread}, \emph{threads per block}, 
\emph{share memory capacity per block}. 
Decreasing these numbers increases parallelism, but 
causing more inter-thread/warp/block communications 
and more launch overhead. The last option can be converted into the option \emph{blocks per SM}.

\textbf{Memory access options.}
We generalize the memory access patterns in Section~\ref{sec:bg} using a combinatorial addressing model.
The address bits $L$ of an $n$-qubit state vector are partitioned into three sets:
$L^{\rm round}$, $L^{\rm batch}$, and $L^{\rm local}$, which index memory access rounds, batches within a round, and amplitudes within a batch, respectively.
Fixing all address bits identifies a single amplitude in the state vector.
A tile is then formed by fixing the values of the bits in $L^{\rm round}$ and enumerating all assignments to the bits in $L^{\rm batch}$ and $L^{\rm local}$.
Within a tile, an amplitude batch is formed by additionally fixing the bits in $L^{\rm batch}$ and varying only those in $L^{\rm local}$.
For a gate $G$ to be tile-local (or batch-local), we hope its qubit index bits belong to $L^{\rm batch} \cup L^{\rm local}$ (or $L^{\rm local}$).  
The addressing design affects read/write locality and data reuse.

We also consider
qubit relayout through index swapping.
This option moves selected high-order qubits to low-order positions
to facilitate amplitude local reuse.


Overall, \cref{tab:qark-kernel-options} provides a parameterization of the PQC kernel.
For each pass $p\in\{\mathrm{fwd},\mathrm{rev}\}$, define
For each forward or reverse pass $p$ of Adiff, a configuration is
\[
  \begin{aligned}
    c = (\mathcal{F},\{c_g\}_{g\in\mathcal{F}}),\ \  
    c_g = (b_g,k_g,\ell_g,a_g,t_g,z_g,\omega_g, \pi).
  \end{aligned}
\]
The fusion partition $\mathcal{F}$ 
specifies the exact gates in each fusion
group; it therefore determines fusion size,
operator structure, and derivative work.  
The full plan is $c=(c^{\mathrm{fwd}},c^{\mathrm{rev}})$; the forward 
and the reverse pass may use different choices. We focus on the reverse pass of Adiff, 
as it dominates the runtime (see Figure \ref{fig:baseline_breakdown}). The forward pass can be likewise reasoned, but without derivative constraints.

\section{PQC Kernel Tuning Framework}
\label{sec:pruning}

The designs in \cref{sec:kdesign} expose a combinatorial configuration
space. \system{} makes tuning tractable by pruning candidates using
resource constraints and reusing search results by exploiting PQC structural similarity.

\subsection{PQC Kernel Resource Model}
\label{sec:pruning-model}

\textbf{For a shared-memory-resident implementation} of Adiff gradient computation, 
if the address assignment keeps required amplitude
interactions tile-local: $b_g
=2^{|L_g^{\mathrm{local}}|+|L_g^{\mathrm{batch}}|}$,
\begin{equation}
\begin{aligned}
b_g
\le
\min\!\left\{
a_gt_g,\,
\frac{
\min\{S_h^{\mathrm{blk}},S_h/z_g\}
-\beta k_g/2-S_g^{\mathrm{other}}
}{
\beta(2+k_g)
}
\right\}.
\end{aligned}
\label{eq:pruning-tile-capacity}
\end{equation}
Here $S_g^{\mathrm{other}}$ is auxiliary shared storage (e.g., gate coefficients) required. $\beta$ is the number of bytes needed for an amplitude.
$\beta k_g/2$ models 
the storage requirement of the real-valued gradient. \textbf{Other notations are explained in Table \ref{tab:qark-kernel-options}}.

Let $S_g$ and $R_g$ denote the resulting shared-memory bytes
and registers allocated per block. Then the residency bound of thread blocks
\begin{equation}
z_g\le
\min\!\left\{
B_h,\,
T_h/t_g,\,
S_h/S_g,\,
R_h/R_g,\, 
N/(b_g H_g)
\right\},
\label{eq:pruning-residency}
\end{equation}
here $N/b_g$ models the number of tile access rounds. 

These constraints provides guidelines for selecting kernel configuration parameters.

\subsection{PQC Kernel Search Space Pruning}
\label{sec:pruning-generation}

We consider reducing the number of candidate kernel configurations to avoid exponential search overhead.
The discussion focuses on Adiff reverse pass. The forward pass can be likewise discussed, but without derivative concerns.

\textbf{Symbolic fixed-support contraction.}
We first contract maximal contiguous sequences of gates with the same
one- or two-qubit support into a composite operator, retaining all
parameter and derivative dependencies. This reduces the number of
fusion boundaries considered during search.

\textbf{Sparse fusion intervals.}
Let $L'=O(n)$ be the number of ordered gates in a $n$-qubit PQC layer, subjecting to fusion search.
We use an edge $i\rightarrow j$ represents fusing gates $i+1,\ldots,j$. 
Candidate lengths are drawn from
\begin{center}
$  \mathcal D
  =
  \{1,\cdots,W\}
  \cup
  \bigl(
    \{2W,4W,8W,\cdots, L'\} \bigr),
$
\end{center}
where $W$ is a bounded short-range search window.
The search overall considers only
$E_{fusion}=O(L'(W+\log L'))=O(n\log n)$ intervals. This sparse fusion strategy may favor amplitude reuse due to the gate locality.

\textbf{Joint resource constraints.}
For each fusion group, the resource model limits the valid selection of parameters like $a_g, k_g$, etc. We also require $b_g > 2^{L_g^{\mathrm{local}}}$. Otherwise, the group $g$ may be too large and should be split; or increase $b_g$. 

\textbf{Address bit mappings.}
For each fusion group, we first identify the required local bits
$L_g^{\mathrm{local}}$.
For tile exponent $d=\log_2 b_g$, we consider two polices for assigning $L_g^{\mathrm{batch}}$: 

(P1) $L_g^{\mathrm{batch}}$ with smallest index bit ids, which favors contiguous accesses,
or control uniformity. 

(P2) $L_g^{\mathrm{batch}}$ with largest index bit ids, which favors more frequent amplitude pairs.

For $L_g^{\mathrm{local}}$, We assign it to prioritize thread-local,
warp-local, and then inter-warp positions, to reduce data movement latency. $L_g^{\mathrm{round}}$ is naturally determined after the other two address lines assigned.

\textbf{Index swapping exploration.}
At the start of each fusion group, 
we determine whether to swap qubits involved in the fusion group to become 
low order ones or just do nothing. Re-layout costs are included when evaluating the plan.

\begin{figure}[t]
  \centering
    \includegraphics[width=0.45\textwidth]{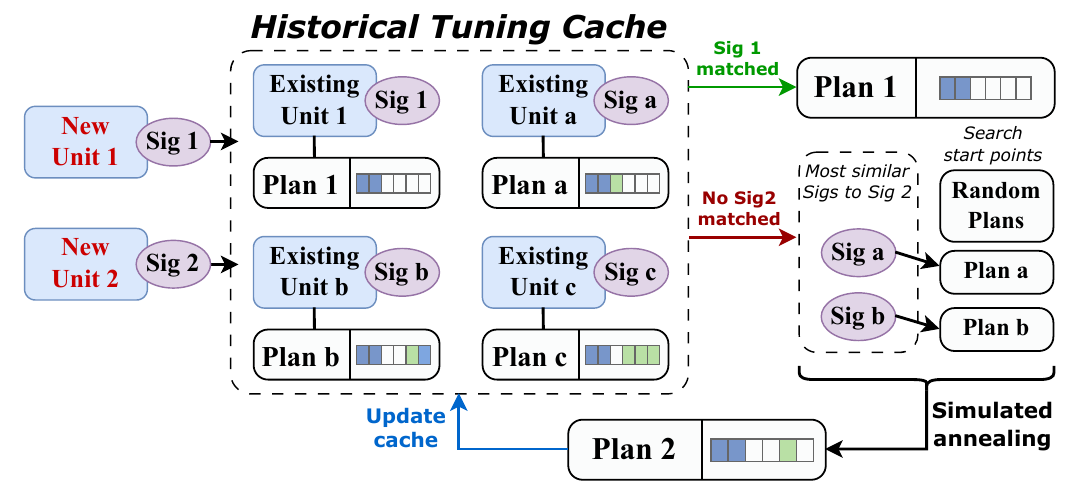}
  \caption{Searching the PQC kernel execution plan based on structural reuse. The tuning unit in the figure is a recurring gate block/layer in the PQC.}
  \label{fig:sa}
\end{figure}

\subsection{Efficient Kernel Search with Structural Reuse}
\label{sec:pruning-search}

PQC layers often repeat, and training typically changes parameter
values while preserving circuit structure. We therefore tune
representative structural units and reuse their plans across
matching occurrences and training iterations,
to form an efficient kernel configuration for the overall PQC.
The tuning cost is amortized, considering PQC training often need to 
run the full circuit thousands of times.

Besides, to enable good search starts, we reuse history search results
as initialization search points. To avoid stuck at local minima, 
we further consider several random initialization points. Figure \ref{fig:sa} summarizes the search process.

\textbf{PQC tuning units.}
We define a tuning unit of a PQC is a gate layer or a short recurring block of layers.
We identify each unit by a structural signature, which is a vector 
records the ordered
gate types and qubit supports, derivative
dependencies, state size, and target GPU. 
The default layout orders address bits as
$q_n,\ldots,q_1$, from most to least significant.
Units with matching signatures can share an execution plan.
Non-matching units with highest cosine similarity
can provide initialization candidates for searching. 
Before tuning, we first pick distinguished tuning units 
from the considered PQC, across training iterations.

\textbf{History-guided and random initialization.} 
For each representative, \system{} uses $S=10$ search starts:
half from historical plans with similar signatures and half from
randomly generated feasible plans.
Historical plans are selected based on the signature similarity between
the history tuning unit and the current unit.
These complementary starting points 
increase the probability of reaching good solutions.

\textbf{Discrete configuration search.}
From each starting point, \system{} runs simulated
annealing~\cite{sa} over the pruned search space.
A search state specifies a fusion partition and its kernel
configurations; neighboring states vary individual discrete choices
while satisfying feasibility constraints.
The objective is the tuning unit's measured execution latency.
We cache evaluated candidates and retain the fastest measured plan
for execution and future search initialization.

Finally, we assemble the tuned units into the full PQC kernel
schedule, inserting qubit permutations where necessary to reconcile
their layouts. The resulting plans are reused across PQC layers and iterations.

%% file: figures/qark_kernel_options.tex
\providecommand{\QarkTuningTableSetup}{%
  \definecolor{QarkTableTeal}{HTML}{247E74}%
  \definecolor{QarkTableTint}{HTML}{EDF5F2}%
  \small
  \renewcommand{\arraystretch}{1.15}%
}
\providecommand{\QarkTuningBand}[2]{%
  \rowcolor{QarkTableTint}%
  \multicolumn{#1}{l}{\textcolor{QarkTableTeal}{#2}}\\%
}

\begin{table*}[t]
  \centering
  \caption{Adiff-aware PQC kernel tuning space and choices retained after pruning.
  Let $n$ denote PQC qubit count, $N=2^n$. $S,R,T,B,H$ denote shared-memory capacity,
  registers, threads, blocks per SM, and SM count; superscripts $\mathrm{blk}$ and $\mathrm{thr}$ denote per-block and per-thread
  limits; Subscript $g$ or $h$ indicates the option is local to a fusion group $g$ or a GPU $h$.}\label{tab:qark-kernel-options} \vspace{-3pt}
  \begingroup
  \QarkTuningTableSetup
  \definecolor{QarkOptionInk}{HTML}{243E50}
  \definecolor{QarkOptionPale}{HTML}{EDF2F5}
  \renewcommand{\arraystretch}{1.15}
  \begin{tabularx}{\linewidth}{
      >{\raggedright\arraybackslash}p{0.165\linewidth}
      >{\centering\arraybackslash}p{0.1\linewidth}
      >{\raggedright\arraybackslash}p{0.28\linewidth}
      >{\raggedright\arraybackslash}X}
    \toprule
    \rowcolor{QarkOptionInk}
    \textcolor{white}{\textbf{Kernel option (\S \ref{sec:ova})}} &
    \textcolor{white}{\textbf{Symbol}} &
    \textcolor{white}{\textbf{Configuration domain}} &
    \cellcolor{QarkTableTeal}%
    \textcolor{white}{\textbf{Retained choices after pruning (\S \ref{sec:pruning-generation})}} \\
    \midrule
    \QarkTuningBand{4}{\textbf{Kernel launch optimization}}
    Fusion partition & $\mathcal{F}$ &
      More than $B_n$ (Bell number) different partition schemes &
      $O(n\log n)$ schemes by fusing gates related to a continuous qubit interval. \\
    Concurrent derivatives & $k_g$ &
      $\forall g \in \mathcal{F}$, $1\le k_g\le g$'s parameter count  &
      Jointly constrained by Equation \ref{eq:pruning-tile-capacity} (\cref{sec:pruning-model}). \\
    \midrule
    \QarkTuningBand{4}{\textbf{GPU parallelism}}
    Amplitudes per thread & $a_g$ &
      $a_g\ge1$. &
       \\
    Threads per block & $t_g$ &
      $1\le t_g\le T_h^{\mathrm{blk}}$. &
      Jointly constrained by Equation \ref{eq:pruning-tile-capacity} (\cref{sec:pruning-model}) \\
    Thread blocks per SM & $z_g$ &
      $z_g\le S_h/S_g$. More in Equation \ref{eq:pruning-residency}. &
       \\
    Amplitudes per tile & $b_g$ &
        $b_g \le a_g t_g$. &
         \\
    \midrule
    \QarkTuningBand{4}{\textbf{Memory access design}}
    \multirow{2}{3cm}{Address mapping $\ell_g$ \hphantom{L}(\cref{sec:ova})} & \multirow{2}{0.5cm}{$L_g^{\mathrm{batch}}$\\[5pt]$L_g^{\mathrm{round}}$} &
      \multirow{2}{3.2cm}{Any subset of non-local bits of size $n-|L_g^{\mathrm{local}}|$.} &
      \textbf{Policies:} (P1) select lowest eligible index bits for $L_g^{\mathrm{batch}}$; (P2) select highest eliglible
      ones for $L_g^{\mathrm{batch}}$. \\
      &  &
         & Remaining non-local bits form $L_g^{\mathrm{round}}$. \\
     & $L_g^{\mathrm{local}}$ & Any permutation of $g$'s qubit list.
       &
       Thread-local, warp-local assignments. \\
    Qubit relayout & $\pi$ &
      $n!$ qubit-index permutations. &
      Keep the layout \textbf{or} move involved qubits
      to low-order positions. Relayout incurs overhead. \\
    \midrule
    \QarkTuningBand{4}{\textbf{Other switches} $\boldsymbol{\omega_g}$ specify gate kernel implementations considered in Pennylane Lightning \cite{pennylane-lightning-docs}.}
    \midrule
  \end{tabularx}
  \endgroup
\end{table*}

%% file: sections/08-implementation.tex





%% file: sections/09-evaluation.tex
\section{Evaluation}\label{sec:evaluation}

In this section, we conduct extensive experiments to evaluate the effectiveness and overhead of \system{}.

\begin{table}[t]
  \caption{Specifications of the evaluated GPUs.}
  \label{tab:platforms}
  \centering
  \small
\begin{tabular}{ccccc}
\hline
GPU Model          & SMs  & L2 Cache & Memory & Shared memory \\ \hline
RTX 6000 Ada & 142 & 96MB    & 48GB  & 100 KB \\
RTX 4090     & 128 & 72MB    & 24GB  & 100 KB \\ \hline
\end{tabular}
\end{table}

\begin{figure}[t]
  \centering
    \includegraphics[width=0.48\textwidth]{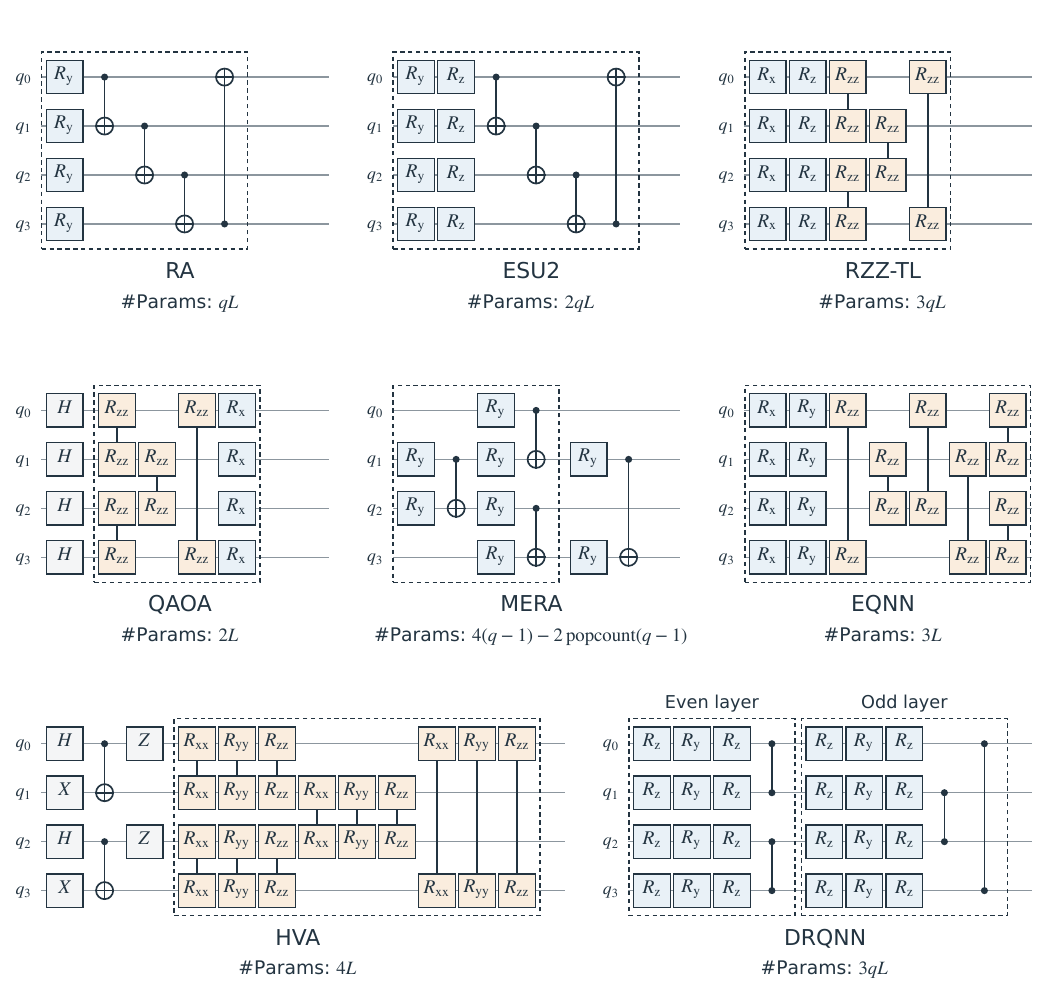}
  \caption{The illustration and the number of independent trainable parameters (\#Params) of each benchmark circuit. The $q$ and $L$ denote the number of qubits and the number of layers in the circuit, respectively. The dashed box denotes the repeated structure.}
  \label{fig:benchmark-circuit}
\end{figure}

\subsection{Experimental Methodology}\label{sec:evaluation-methodology}

{\bf Platforms.} We conduct the evaluations on two servers with GPUs of different hardware specifications.
Each server is equipped with two 2.45GHz AMD EPYC 7763 CPUs and 128GB of DDR4 memory.
One server is equipped with an NVIDIA RTX 4090 GPU, while the other is equipped with an NVIDIA RTX 6000 Ada GPU. 
\Cref{tab:platforms} summarizes the specifications of the evaluated GPUs.
The servers run Ubuntu 24.04.4 with NVIDIA driver~595.84, CUDA~13.2, Python~3.12.3, and NumPy~2.5.1. 
We compile \system{} using \texttt{nvcc -O3 -std=c\\++17 -arch=native}.
Consistent with prior work~\cite{haner20175, guerreschi2020intel, suzuki2021qulacs}, we store both the real and imaginary components of the state vector's amplitude in FP64.

\begin{figure*}[t]
  \centering
  \includegraphics[width=0.95\textwidth]{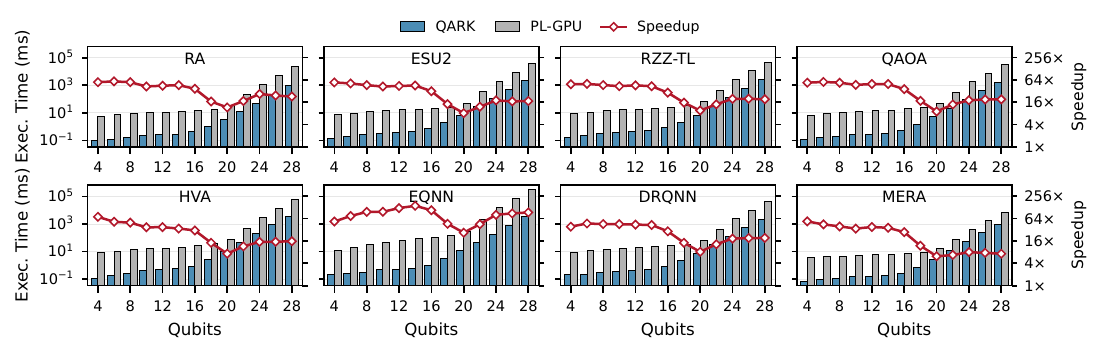}
  \caption{Comparison of \system{} with \baseline{} (PennyLane-Lightning-GPU simulator) on RTX 6000 Ada in terms of Adiff computation execution time. The speedup is computed by ‘Execution time of \baseline{} / Execution time of \system{}’.}
  \label{fig:main-circuits}
\end{figure*}

\begin{figure*}[t]
  \centering
  \includegraphics[width=0.94\textwidth]{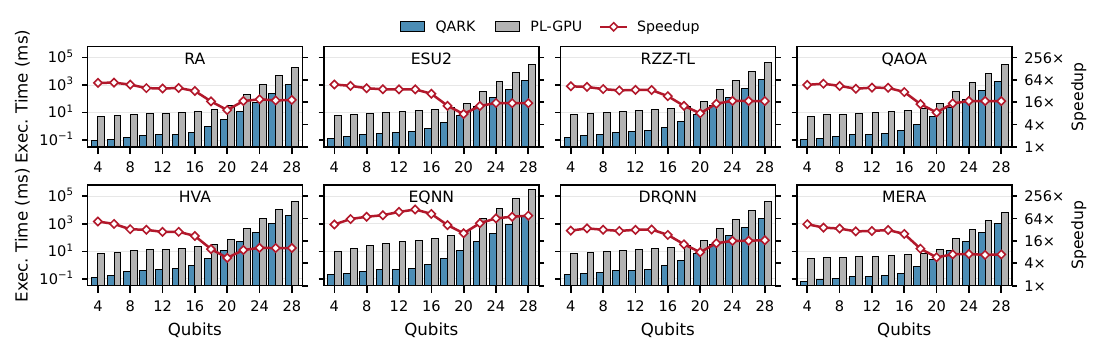}
  \caption{Comparison of \system{} with \baseline{} (PennyLane-Lightning-GPU simulator) on RTX 4090 in terms of Adiff computation execution time. The speedup is computed by ‘Execution time of \baseline{} / Execution time of \system{}’.}
  \label{fig:platform-circuits}
\end{figure*}



{\bf Baseline.} We compare \system{} with the state-vector simulator \texttt{lightning.gpu}~0.45.0 of PennyLane~\cite{pennylane-lightning-docs}, which supports Adiff~\cite{asadi2024, pennylane-support-adjoint}.
We hereafter refer to this baseline as ``\baseline{}''.


{\bf Benchmark Circuits.} We use eight PQCs with 4, 6, 8, ..., 26, and 28 qubits in our evaluation.
The data for circuits with more qubits cannot fit entirely in the memory of the evaluated GPUs.
The used circuits include three Hardware-Efficient Ansatz (HEA) circuits from the Qiskit Circuit Library~\cite{qiskit-circuit-library}---RealAmplitudes (RA), EfficientSU2 (ESU2) and RZZ-entangled TwoLocal (RZZ-TL), Quantum Approximate Optimization Algorithm (QAOA) on MAXCUT problems~\cite{qaoa},  XXZ Hamiltonian Variational Ansatz (HVA)~\cite{PRXQuantum.1.020319}, Equivariant Quantum Neural Network (EQNN)~\cite{PRXQuantum.4.010328,Schatzki2024}, Data Re-uploading Quantum Neural Network (DRQNN)~\cite{perez2020data}, and Multi-scale Entanglement Renormalization Ansatz (MERA)\\~\cite{grant2018hierarchical}.
These circuits are widely used in prior work on quantum circuit simulation~\cite{lykov2021performance,efthymiou2022qibo,suzuki2021qulacs,efthymiou2022quantum,tensorcircuit2023,zhang2025bmqsim}.

Figure~\ref{fig:benchmark-circuit} illustrates the structure and the number of independent trainable parameters of each circuit.
Except for MERA, the other circuits can be constructed by stacking an arbitrary number $L$ of repeated structures to form an $L$-layer circuit.
For a multi-layer MERA circuit with $q$ qubits, the number of qubits participating in the R$_y$ gates is halved at each subsequent layer. Consequently, the number of layers and the number of trainable parameters are not independently configurable, but are determined by $q$. The number of layers is $\lceil \log_2 q \rceil$.
The number of trainable parameters is given in the figure; here, \texttt{popcount} denotes the number of bits set to 1 in the binary representation of an integer.
We set the circuit depth to eight layers in the following experiments and evaluate the impact of circuit depth in the sensitivity study in Section~\ref{sec:additional-evaluation}.

\subsection{Comparison with PennyLane Lightning}\label{sec:evaluation-end-to-end}

We perform 1000 Adiff computations for each PQC and report the average end-to-end execution time.
The execution time is measured from the CPU launching the first GPU kernel to the final result becoming available to the CPU.
For the eight circuits, we evaluate 13 qubit-count configurations per circuit, yielding a total of 8 $\times$ 13 = 104 test cases.

{\bf Speedups across circuits and GPUs.}
\Cref{fig:main-circuits} reports the execution times when using \system{} and \baseline{} on the RTX 6000 Ada GPU.
The red line on the secondary y-axis shows the speedup of \system{} over \baseline{};
as shown, \system{} outperforms \baseline{} across all test cases, achieving an overall geometric-mean speedup of $29.4\times$ and a maximum speedup of $141.8\times$ on 14-qubit EQNN.

\system{} fundamentally improves the practicality of PQC training simulation.
As discussed in Section~\ref{sec:PL-vs-Qiskit}, \baseline{} cannot provide sufficient performance for efficient PQC training simulation. 
Compared with \baseline{}, QARK reduces the execution time per Adiff from 2s to 45.2ms for a 22-qubit EQNN and from 46.18ms to 0.47ms for a 10-qubit EQNN. 
Consequently, the overall training time decreases from 5.56 hours to 7.5 minutes for the 22-qubit EQNN and from 7.7 minutes to 4.7 seconds for the 10-qubit EQNN.


As shown in \Cref{fig:platform-circuits}, the performance advantage of \system{} also holds on the RTX 4090 GPU. 
\system{} is faster than \baseline{} in every test case, with a geometric-mean speedup of $24.7\times$ and a maximum speedup of $113.0\times$ on 14-qubit EQNN.

{\it In summary, \system{} consistently delivers substantial speedups across different GPUs and circuit types.}

{\bf Understanding \system{}'s speedups.}
The speedups result from \system{} identifying and applying appropriate Adiff computation modes for different circuits, whereas \baseline{} uses a fixed mode across all circuits.
With QARK, the GPU can be utilized more effectively.

We examine the GPU kernel configurations found by \system{} in our evaluation and observe that they often vary with the circuit type, qubit count and GPU. Table~\ref{tab:reconfigurable-example} illustrates this variation using QAOA and RA as examples.

\begin{figure}[t]
  \centering
  \includegraphics[width=0.45\textwidth]{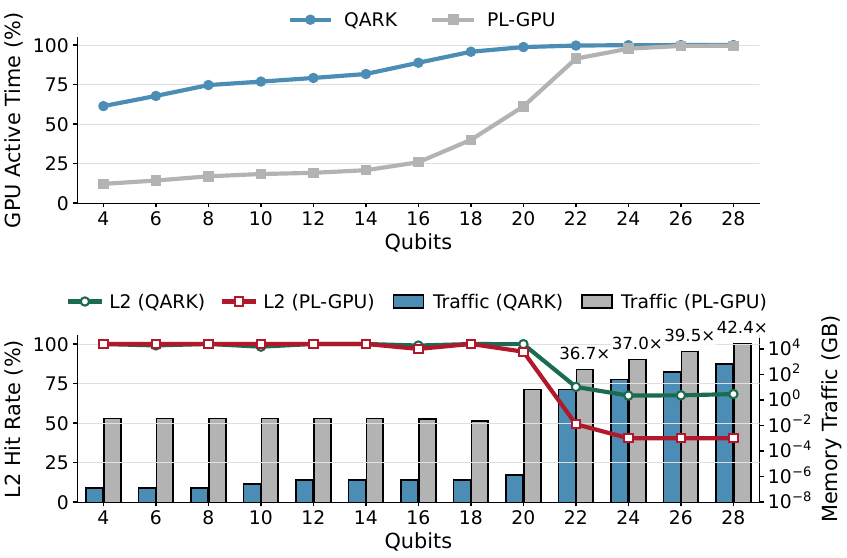}
  \caption{When performing QAOA's Adiff, the top part of the figure shows the GPU active time ratio, while the bottom part shows the L2 cache hit rate and GPU memory traffic.}
  \label{fig:latency-regimes}
\end{figure}

The dedicated GPU kernel configurations provided by \system{} enable more effective GPU utilization.
Figure~\ref{fig:latency-regimes} shows the GPU performance metrics during QAOA's Adiff with varying qubit counts on RTX 6000 Ada GPU, including the ratio of GPU active time, GPU L2 cache hit rate, and total GPU memory traffic (i.e., the total amount of data read from and written to GPU memory during an Adiff).

As observed, \system{} consistently achieves lower kernel-launch and synchronization overheads (indicated by the GPU active time ratio), higher L2 cache hit rates, and less data transferred to/from the GPU memory.
This is why \system{} outperforms \baseline{} on Adiff for QAOA. 
Similar results are also observed on the other circuits.

Note that \system{}'s speedup over \baseline{} decreases as the qubit count increases, reaching its minimum at 20 qubits, where \system{} still achieves at least a 5.6$\times$ speedup. Beyond 20 qubits, the speedup increases with the qubit count. 
Figure~\ref{fig:latency-regimes} explains this trend. 
Below 20 qubits, both \system{} and \baseline{} read/write data at L2 cache, keeping GPU memory traffic negligible (less than 35MB), while \system{}'s advantage in GPU active time ratio narrows with the qubit count. 
Beyond 20 qubits, their GPU active time ratios become similar, but \baseline{}'s lower L2 cache hit rate leads to more memory traffic. This traffic gap widens as the qubit count increases.

\begin{table}[t]
\caption{GPU kernel configurations for QAOA and RA, including amplitudes per thread (APT), threads per block (TPB), share memory capacity per block (CPB), and the selected address bit mapping policy.}
\label{tab:reconfigurable-example}
\centering
\scriptsize
\begin{tabular}{c|c|c|cc|cc|c|c}
\hline
\multirow{2}{*}{\textbf{GPU}} & \multirow{2}{*}{\textbf{Circuit}} & \multirow{2}{*}{\textbf{\#Qubits}} & \multicolumn{2}{c|}{\textbf{Forward}}            & \multicolumn{2}{c|}{\textbf{Backward}}           & \multirow{2}{*}{\textbf{CPB}} & \multirow{2}{*}{\textbf{Policy}} \\ \cline{4-7}
                              &                                   &                                    & \multicolumn{1}{c|}{\textbf{APT}} & \textbf{TPB} & \multicolumn{1}{c|}{\textbf{APT}} & \textbf{TPB} &                               &                                  \\ \hline
6000                          & QAOA                              & 14                                 & \multicolumn{1}{c|}{4}            & 32           & \multicolumn{1}{c|}{4}            & 32           & 50\%                          & (P1)                             \\ \hline
6000                          & QAOA                              & 22                                 & \multicolumn{1}{c|}{8}            & 128          & \multicolumn{1}{c|}{8}            & 128          & 50\%                          & (P2)                             \\ \hline
6000                          & QAOA                              & 28                                 & \multicolumn{1}{c|}{16}           & 128          & \multicolumn{1}{c|}{16}           & 128          & 50\%                          & (P2)                             \\ \hline
6000                          & RA                                & 14                                 & \multicolumn{1}{c|}{4}            & 32           & \multicolumn{1}{c|}{4}            & 32           & 100\%                         & (P1)                             \\ \hline
6000                          & RA                                & 22                                 & \multicolumn{1}{c|}{4}            & 64           & \multicolumn{1}{c|}{4}            & 64           & 50\%                          & (P2)                             \\ \hline
6000                          & RA                                & 28                                 & \multicolumn{1}{c|}{16}           & 64           & \multicolumn{1}{c|}{16}           & 64           & 50\%                          & (P2)                             \\ \hline
4090                          & QAOA                              & 14                                 & \multicolumn{1}{c|}{4}            & 32           & \multicolumn{1}{c|}{4}            & 32           & 100\%                         & (P1)                             \\ \hline
4090                          & QAOA                              & 22                                 & \multicolumn{1}{c|}{8}            & 128          & \multicolumn{1}{c|}{8}            & 32           & 50\%                          & (P2)                             \\ \hline
4090                          & QAOA                              & 28                                 & \multicolumn{1}{c|}{8}            & 128          & \multicolumn{1}{c|}{8}            & 32           & 50\%                          & (P2)                             \\ \hline
4090                          & RA                                & 14                                 & \multicolumn{1}{c|}{4}            & 32           & \multicolumn{1}{c|}{4}            & 32           & 50\%                          & (P1)                             \\ \hline
4090                          & RA                                & 22                                 & \multicolumn{1}{c|}{8}            & 64           & \multicolumn{1}{c|}{8}            & 64           & 100\%                         & (P2)                             \\ \hline
4090                          & RA                                & 28                                 & \multicolumn{1}{c|}{16}           & 128          & \multicolumn{1}{c|}{16}           & 128          & 100\%                         & (P2)                             \\ \hline
\end{tabular}
\end{table}

\subsection{Additional Study}\label{sec:additional-evaluation}

We conduct ablation and sensitivity studies to better understand the design space of \system{}, and also analyze its overhead. 
As the observations and conclusions are consistent across the RTX 6000 Ada and RTX 4090 GPUs, we report only the results of former in this section.

\begin{figure}[t]
  \centering
  \includegraphics[width=0.49\textwidth]{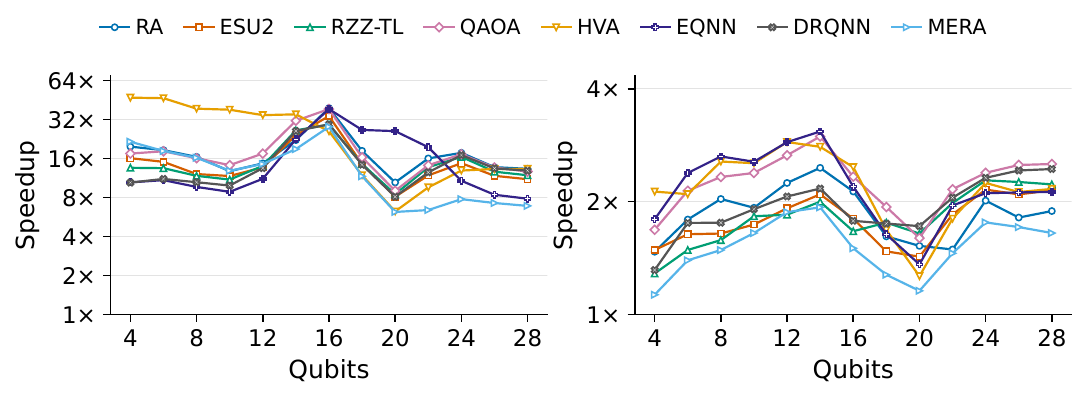}
  \caption{\system{}'s speedup over \system{}-NP (left part of the figure) and \system{}-NS (right part).}
  \label{fig:ablation_study}
\end{figure}

{\bf Effectiveness the kernel tuning framework in \system{}.}
We conduct experiments to independently evaluate the effectiveness of specific designs in \system{}.
Two variations of \system{} are implemented: (1) {\bf \system{}-NP.} When finding a kernel for a circuit, we disable search-space pruning and instead randomly sample kernel configurations from the original large search space to perform Adiff computation, reporting their average execution time. (2) {\bf \system{}-NS.} After search-space pruning, we do not conduct any further kernel search; instead, we randomly sample kernel configurations from the pruned space to perform Adiff computation and report their average execution time. 

Figure~\ref{fig:ablation_study} shows QARK's speedups over the two variants. 
For \system{}-NP, \system{} achieves a geometric-mean speedup of 15.2$\times$, with a maximum speedup of 47.2$\times$. For \system{}-NS, \system{} achieves a geometric-mean speedup of 1.9$\times$ and up to 3$\times$. These results demonstrate the necessity and effectiveness of \system{}'s kernel tuning framework. 

\begin{figure}[t]
  \centering
  \includegraphics[width=0.49\textwidth]{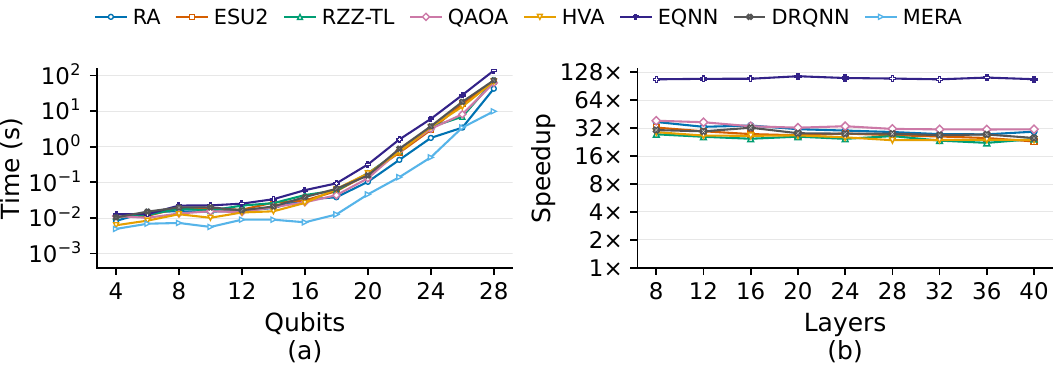}
  \caption{(a) The time required by \system{} to find a kernel. (b) \system{}'s speedup over \baseline{} as circuit depth increases. Each circuit is configured to 16 qubits.}
  \label{fig:search_time_and_layer}
\end{figure}

{\bf Time spent on kernel search.}
We measure the time required for \system{} to find an ideal kernel configuration within the pruned search space, as reported in Figure~\ref{fig:search_time_and_layer}(a).
The search time increases with the number of qubits, reaching up to 136.6s for 28-qubit EQNN.
Such overhead is acceptable because, for a given circuit and GPU, the search only needs to be done once offline.

{\bf Sensitivity to circuit depth.}
We evaluate \system{}'s speedup over \baseline{} when we increase the number of PQC layers from 8 to 12, 16, ..., 36, and 40, while fixing the PQC at 16 qubits.
\system{} uses the same GPU kernel configuration for each circuit as in Section~\ref{sec:evaluation-end-to-end}.
As shown in Figure~\ref{fig:search_time_and_layer}(b), the speedups remain roughly unchanged as the circuit depth increases, indicating that the GPU kernel configurations selected by \system{} based on partial shallow circuits remain effective for whole deeper circuits.

\begin{figure}[t]
\centering
\includegraphics[width=0.4\textwidth]{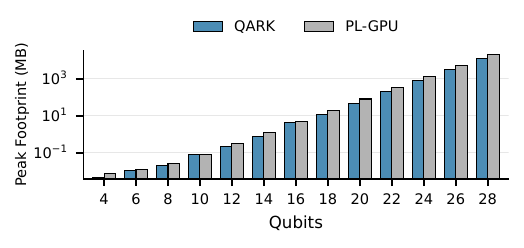}
\caption{The peak memory footprint during Adiff computation of ESU2.}
\label{fig:memory-footprint}
\end{figure}

{\bf GPU memory footprint.} 
We measure the peak GPU memory footprint during Adiff computation for both \system{} and \baseline{}. 
Among the eight evaluated circuits, ESU2 incurs the largest footprint when using QARK; Figure~\ref{fig:memory-footprint} reports its peak memory footprint. 
As shown, \system{} does not require more GPU memory capacity than \baseline{}, indicating that its speedups are not achieved at the cost of increased memory capacity.

%% file: sections/10-related-work.tex
\section{Related Work}
\label{sec:related-work}

{\bf Simulators for general quantum circuits.} 
Quantum circuit simulation plays an important role in developing quantum algorithms, and many prior studies~\cite{deraedt2007simulator,haner20175,jones2019quest,suzuki2021qulacs,cuquantum2023,qgpu2022,svsim2021,uniq2022,doi2020cache,hyquas2021,atlas2024,bqsim2025,qblaze2025,cast2025,dagfusion2026,qfusion2026} focus on developing and optimizing general-purpose quantum circuit simulators.
QuEST~\cite{jones2019quest}, Qulacs~\cite{suzuki2021qulacs}, UniQ~\cite{uniq2022}, cuQuantum~\cite{cuquantum2023} and CAST~\cite{cast2025} aim to provide reusable primitives and execution backends for diverse quantum circuits across multiple hardware platforms.
BQSim~\cite{bqsim2025}, qfusion-opt~\cite{qfusion2026}, HyQuas~\cite{hyquas2021} and ~\cite{dagfusion2026} improve simulation by scheduling/fusing gates and batching input state-vectors.
SV-Sim~\cite{svsim2021}, Atlas~\cite{atlas2024} and ~\cite{haner20175} focus on developing distributed state-vector simulators to scale circuit execution through communication-aware scheduling, qubit remapping, and efficient inter-device data movement.
BMQSim~\cite{zhang2025bmqsim} reduces state-vector memory requirements to support larger circuits while controlling approximation error.
While \system{} can incorporate above optimizations, \system{} specifically targets PQC training simulation rather than general simulation. 
Targeting PQC training is necessary because general-purpose simulators can take several hours to complete it, as they are not designed for the gradient computation required specifically by PQC training, which dominates the simulation time.

{\bf Simulators with dedicated PQC training support.}
While TensorCircuit~\cite{tensorcircuit2023} supports PQC simulation based on tensor-network, most prior work~\cite{luo2020yao,tfq2020, asadi2024,qandle2025,tqml2025,julivqc2024,catalyst2024,vqcsim2026} on PQCs is based on state-vector simulation.
PQC's gradients can be computed in multiple ways.
Catalyst~\cite{cast2025} and ~\cite{schuld2019parameter} adopt the PSR method, while Yao~\cite{luo2020yao}, VQCSim~\cite{vqcsim2026}, TQml~\cite{tqml2025} and Qandle~\cite{qandle2025} use the BP method.
QARK focuses on Adiff because it combines the advantages of the above two methods and has been adopted in recent studies and widely-used QML frameworks such as PennyLane and TensorFlow Quantum (TFQ)~\cite{tfq2020, julivqc2024,asadi2024}.
While only PennyLane provides a native GPU-based simulator, its fixed GPU kernel schedule prevents it from achieving ideal performance. Instead, QARK exploits the optimization opportunities enabled by reconfigurable GPU kernels.

{\bf Customizing execution for circuits.}
Customizing the execution to workload characteristics or hardware specifications is essential for achieving further performance improvements.
Inspired by prior studies~\cite{tvm2018,ansor2020,enzyme2021,roller2022} on automatically generating and tuning GPU kernels for deep learning workloads, \system{} proposes a set of differentiation-aware GPU kernel designs together with the efficient autotuning method according to PQCs' characteristics.
QOKit~\cite{qokit2023} and CUAOA~\cite{stein2024cuaoa} design GPU kernels specifically for QAOA, whereas QARK is more general and adaptively selects kernels for a broader range of PQCs with circuit structures different from QAOA.

%% file: sections/12-conclusion.tex

\section{Discussion and Future Work}

\system demonstrates the importance of reconfigurability for efficient  PQC simulation. High performance depends not only on the available kernel optimizations, but also on how they are configured for a particular workload and GPU architecture. By exposing these choices and combining resource-guided pruning with structural reuse, \system provides an efficient and low-overhead mechanism for retuning kernels as workload characteristics or GPU resources change. This flexibility helps maintain high performance across the evaluated configurations. Although \system already substantially outperforms the baseline across various workload-hardware combinations, opportunities remain to reduce adaptation costs and extend its applicability:

1) Further reducing tuning overhead would make \system more effective for workloads with limited opportunities for reuse. The current design amortizes empirical search across recurring circuit structures and training iterations. For short runs or frequently changing circuit structures, though seldom encountered, the search cost may be harder to amortize. Building on the existing history-guided initialization, future work could use predictive cost models to prioritize promising configurations and adapt search budgets to the expected number of subsequent executions. Optimizing the combined cost of tuning and execution would help \system respond more efficiently to workload changes.

2) Scaling \system beyond a single GPU would expand the range of PQCs it can simulate. The current implementation retains the exponential storage requirement of dense state-vector simulation, limiting circuit size to the memory capacity of one device. A multi-GPU extension would need to coordinate state partitioning, qubit placement, and gradient accumulation while accounting for communication during both forward execution and the reverse pass. Incorporating these costs into the resource model and tuning objective could help identify configurations that balance local computation with inter-device data movement.

\section{Conclusion}
\label{sec:conclusion}

We presented \system, a reconfigurable GPU kernel framework for accelerating PQC training simulation. \system is motivated by two observations: (a) existing general-purpose simulators lack native support for backward gradient computation in PQCs; and (b) simulation performance depends strongly on both the underlying GPU architecture and the PQC workload. Overall, \system identifies and addresses three challenges: adapting general circuit simulation optimizations to gradient computation, making the combinatorial configuration space for a given workload and hardware combination tractable, and amortizing empirical search across repeated circuit structures and training iterations. Consequently, \system achieves efficient PQC training simulation, regardless of the workload-hardware combination. Moreover, \system incurs only modest tuning overhead by exploiting PQC's structural similarities across circuit layers and training iterations.

%% file: main.bbl

\begin{thebibliography}{59}


\ifx \showCODEN    \undefined \def \showCODEN     #1{\unskip}     \fi
\ifx \showDOI      \undefined \def \showDOI       #1{#1}\fi
\ifx \showISBNx    \undefined \def \showISBNx     #1{\unskip}     \fi
\ifx \showISBNxiii \undefined \def \showISBNxiii  #1{\unskip}     \fi
\ifx \showISSN     \undefined \def \showISSN      #1{\unskip}     \fi
\ifx \showLCCN     \undefined \def \showLCCN      #1{\unskip}     \fi
\ifx \shownote     \undefined \def \shownote      #1{#1}          \fi
\ifx \showarticletitle \undefined \def \showarticletitle #1{#1}   \fi
\ifx \showURL      \undefined \def \showURL       {\relax}        \fi
\providecommand\bibfield[2]{#2}
\providecommand\bibinfo[2]{#2}
\providecommand\natexlab[1]{#1}
\providecommand\showeprint[2][]{arXiv:#2}

\bibitem[Asadi et~al\mbox{.}(2024)]%
        {asadi2024}
\bibfield{author}{\bibinfo{person}{Ali Asadi}, \bibinfo{person}{Amintor Dusko},
  \bibinfo{person}{Chae-Yeun Park}, \bibinfo{person}{Vincent Michaud-Rioux},
  \bibinfo{person}{Isidor Schoch}, \bibinfo{person}{Shuli Shu},
  \bibinfo{person}{Trevor Vincent}, {and} \bibinfo{person}{Lee~James
  O'Riordan}.} \bibinfo{year}{2024}\natexlab{}.
\newblock \bibinfo{title}{Hybrid Quantum Programming with {PennyLane Lightning}
  on {HPC} Platforms}.
\newblock
\newblock
\urldef\tempurl%
\url{https://doi.org/10.48550/arXiv.2403.02512}
\showDOI{\tempurl}
\showeprint[arxiv]{2403.02512}~[quant-ph]


\bibitem[Bayraktar et~al\mbox{.}(2023)]%
        {cuquantum2023}
\bibfield{author}{\bibinfo{person}{Harun Bayraktar}, \bibinfo{person}{Ali
  Charara}, \bibinfo{person}{David Clark}, \bibinfo{person}{Saul Cohen},
  \bibinfo{person}{Timothy~B. Costa}, \bibinfo{person}{Yao-Lung~L. Fang},
  \bibinfo{person}{Yang Gao}, \bibinfo{person}{Jack Guan},
  \bibinfo{person}{John~A. Gunnels}, \bibinfo{person}{Azzam Haidar},
  \bibinfo{person}{Andreas Hehn}, \bibinfo{person}{Markus H{\"o}hnerbach},
  \bibinfo{person}{Matthew Jones}, \bibinfo{person}{Tom Lubowe},
  \bibinfo{person}{Dmitry~I. Lyakh}, \bibinfo{person}{Shinya Morino},
  \bibinfo{person}{Paul Springer}, \bibinfo{person}{Sam Stanwyck},
  \bibinfo{person}{Igor Terentyev}, \bibinfo{person}{Satya Varadhan},
  \bibinfo{person}{Jonathan Wong}, {and} \bibinfo{person}{Takuma Yamaguchi}.}
  \bibinfo{year}{2023}\natexlab{}.
\newblock \showarticletitle{{cuQuantum SDK}: A High-Performance Library for
  Accelerating Quantum Science}. In \bibinfo{booktitle}{\emph{IEEE
  International Conference on Quantum Computing and Engineering}}.
  \bibinfo{publisher}{IEEE}, \bibinfo{address}{Piscataway, NJ, USA},
  \bibinfo{pages}{1050--1061}.
\newblock
\urldef\tempurl%
\url{https://doi.org/10.1109/QCE57702.2023.00119}
\showDOI{\tempurl}


\bibitem[Broughton et~al\mbox{.}(2020)]%
        {tfq2020}
\bibfield{author}{\bibinfo{person}{Michael Broughton},
  \bibinfo{person}{Guillaume Verdon}, \bibinfo{person}{Trevor McCourt},
  \bibinfo{person}{Antonio~J. Martinez}, \bibinfo{person}{Jae~Hyeon Yoo},
  \bibinfo{person}{Sergei~V. Isakov}, \bibinfo{person}{Philip Massey},
  \bibinfo{person}{Ramin Halavati}, \bibinfo{person}{Murphy~Yuezhen Niu},
  \bibinfo{person}{Alexander Zlokapa}, \bibinfo{person}{Evan Peters},
  \bibinfo{person}{Owen Lockwood}, \bibinfo{person}{Andrea Skolik},
  \bibinfo{person}{Sofiene Jerbi}, \bibinfo{person}{Vedran Dunjko},
  \bibinfo{person}{Martin Leib}, \bibinfo{person}{Michael Streif},
  \bibinfo{person}{David~Von Dollen}, \bibinfo{person}{Hongxiang Chen},
  \bibinfo{person}{Shuxiang Cao}, \bibinfo{person}{Roeland Wiersema},
  \bibinfo{person}{Hsin-Yuan Huang}, \bibinfo{person}{Jarrod~R. McClean},
  \bibinfo{person}{Ryan Babbush}, \bibinfo{person}{Sergio Boixo},
  \bibinfo{person}{Dave Bacon}, \bibinfo{person}{Alan~K. Ho},
  \bibinfo{person}{Hartmut Neven}, {and} \bibinfo{person}{Masoud Mohseni}.}
  \bibinfo{year}{2020}\natexlab{}.
\newblock \bibinfo{title}{{TensorFlow Quantum}: A Software Framework for
  Quantum Machine Learning}.
\newblock
\newblock
\urldef\tempurl%
\url{https://doi.org/10.48550/arXiv.2003.02989}
\showDOI{\tempurl}
\showeprint[arxiv]{2003.02989}~[quant-ph]


\bibitem[Chen et~al\mbox{.}(2018)]%
        {tvm2018}
\bibfield{author}{\bibinfo{person}{Tianqi Chen}, \bibinfo{person}{Thierry
  Moreau}, \bibinfo{person}{Ziheng Jiang}, \bibinfo{person}{Lianmin Zheng},
  \bibinfo{person}{Eddie Yan}, \bibinfo{person}{Haichen Shen},
  \bibinfo{person}{Meghan Cowan}, \bibinfo{person}{Leyuan Wang},
  \bibinfo{person}{Yuwei Hu}, \bibinfo{person}{Luis Ceze},
  \bibinfo{person}{Carlos Guestrin}, {and} \bibinfo{person}{Arvind
  Krishnamurthy}.} \bibinfo{year}{2018}\natexlab{}.
\newblock \showarticletitle{{TVM}: An Automated End-to-End Optimizing Compiler
  for Deep Learning}. In \bibinfo{booktitle}{\emph{USENIX Symposium on
  Operating Systems Design and Implementation}}. \bibinfo{publisher}{USENIX
  Association}, \bibinfo{address}{Carlsbad, CA, USA},
  \bibinfo{pages}{578--594}.
\newblock
\urldef\tempurl%
\url{https://www.usenix.org/conference/osdi18/presentation/chen}
\showURL{%
\tempurl}


\bibitem[Developer(2026)]%
        {pl-gpu-source-code}
\bibfield{author}{\bibinfo{person}{PennlyLane Developer}.}
  \bibinfo{year}{2026}\natexlab{}.
\newblock \bibinfo{title}{AdjointJacobianGPU.hpp}.
\newblock
\newblock
\urldef\tempurl%
\url{https://github.com/PennyLaneAI/pennylane-lightning/blob/main/pennylane_lightning/core/simulators/lightning_gpu/algorithms/AdjointJacobianGPU.hpp}
\showURL{%
\tempurl}


\bibitem[Doi and Horii(2020)]%
        {doi2020cache}
\bibfield{author}{\bibinfo{person}{Jun Doi} {and} \bibinfo{person}{Hiroshi
  Horii}.} \bibinfo{year}{2020}\natexlab{}.
\newblock \showarticletitle{Cache Blocking Technique to Large Scale Quantum
  Computing Simulation on Supercomputers}. In \bibinfo{booktitle}{\emph{2020
  IEEE International Conference on Quantum Computing and Engineering (QCE)}}.
  \bibinfo{publisher}{IEEE}, \bibinfo{address}{Piscataway, NJ, USA},
  \bibinfo{pages}{212--222}.
\newblock
\urldef\tempurl%
\url{https://doi.org/10.1109/qce49297.2020.00035}
\showDOI{\tempurl}


\bibitem[Efthymiou et~al\mbox{.}(2022a)]%
        {efthymiou2022quantum}
\bibfield{author}{\bibinfo{person}{Stavros Efthymiou}, \bibinfo{person}{Marco
  Lazzarin}, \bibinfo{person}{Andrea Pasquale}, {and} \bibinfo{person}{Stefano
  Carrazza}.} \bibinfo{year}{2022}\natexlab{a}.
\newblock \showarticletitle{Quantum simulation with just-in-time compilation}.
\newblock \bibinfo{journal}{\emph{Quantum}}  \bibinfo{volume}{6}
  (\bibinfo{year}{2022}), \bibinfo{pages}{814}.
\newblock


\bibitem[Efthymiou et~al\mbox{.}(2022b)]%
        {efthymiou2022qibo}
\bibfield{author}{\bibinfo{person}{Stavros Efthymiou}, \bibinfo{person}{Sergi
  Ramos-Calderer}, \bibinfo{person}{Carlos Bravo-Prieto},
  \bibinfo{person}{Adri{\'a}n P{\'e}rez-Salinas}, \bibinfo{person}{Diego
  Garc{\'\i}a-Mart{\'\i}n}, \bibinfo{person}{Artur Garcia-Saez},
  \bibinfo{person}{Jos{\'e}~Ignacio Latorre}, {and} \bibinfo{person}{Stefano
  Carrazza}.} \bibinfo{year}{2022}\natexlab{b}.
\newblock \showarticletitle{Qibo: a framework for quantum simulation with
  hardware acceleration}.
\newblock \bibinfo{journal}{\emph{Quantum Science \& Technology}}
  \bibinfo{volume}{7}, \bibinfo{number}{1} (\bibinfo{year}{2022}),
  \bibinfo{pages}{015018}.
\newblock


\bibitem[Farhi et~al\mbox{.}(2014)]%
        {qaoa}
\bibfield{author}{\bibinfo{person}{Edward Farhi}, \bibinfo{person}{Jeffrey
  Goldstone}, {and} \bibinfo{person}{Sam Gutmann}.}
  \bibinfo{year}{2014}\natexlab{}.
\newblock \bibinfo{title}{A Quantum Approximate Optimization Algorithm}.
\newblock
\newblock
\urldef\tempurl%
\url{https://doi.org/10.48550/arXiv.1411.4028}
\showDOI{\tempurl}
\showeprint[arxiv]{1411.4028}~[quant-ph]


\bibitem[Firc et~al\mbox{.}(2026)]%
        {vqcsim2026}
\bibfield{author}{\bibinfo{person}{Anton Firc}, \bibinfo{person}{Martin
  Pere{\v{s}}{\'i}ni}, \bibinfo{person}{Vojt{\v{e}}ch Mr{\'a}zek},
  \bibinfo{person}{Kamil Malinka}, \bibinfo{person}{Vojt{\v{e}}ch
  Stan{\v{e}}k}, \bibinfo{person}{Zbyn{\v{e}}k Li{\v{c}}ka},
  \bibinfo{person}{Nouhaila Innan}, \bibinfo{person}{Walid~El Maouaki},
  \bibinfo{person}{Alberto Marchisio}, {and} \bibinfo{person}{Muhammad
  Shafique}.} \bibinfo{year}{2026}\natexlab{}.
\newblock \bibinfo{title}{{VQCSim}: When Does Compile-Once Statevector
  Simulation Beat Generic Quantum Frameworks?}
\newblock
\newblock
\urldef\tempurl%
\url{https://doi.org/10.48550/arXiv.2607.11985}
\showDOI{\tempurl}
\showeprint[arxiv]{2607.11985}~[quant-ph]
\newblock
\shownote{Accepted at IEEE/ACM International Conference on Computer-Aided
  Design (ICCAD 2026); preprint version}.


\bibitem[Grant et~al\mbox{.}(2018)]%
        {grant2018hierarchical}
\bibfield{author}{\bibinfo{person}{Edward Grant}, \bibinfo{person}{Marcello
  Benedetti}, \bibinfo{person}{Shuxiang Cao}, \bibinfo{person}{Andrew Hallam},
  \bibinfo{person}{Joshua Lockhart}, \bibinfo{person}{Vid Stojevic},
  \bibinfo{person}{Andrew~G Green}, {and} \bibinfo{person}{Simone Severini}.}
  \bibinfo{year}{2018}\natexlab{}.
\newblock \showarticletitle{Hierarchical quantum classifiers}.
\newblock \bibinfo{journal}{\emph{npj Quantum Information}}
  \bibinfo{volume}{4}, \bibinfo{number}{1} (\bibinfo{year}{2018}),
  \bibinfo{pages}{65}.
\newblock


\bibitem[Guerreschi et~al\mbox{.}(2020)]%
        {guerreschi2020intel}
\bibfield{author}{\bibinfo{person}{Gian~Giacomo Guerreschi},
  \bibinfo{person}{Justin Hogaboam}, \bibinfo{person}{Fabio Baruffa}, {and}
  \bibinfo{person}{Nicolas~PD Sawaya}.} \bibinfo{year}{2020}\natexlab{}.
\newblock \showarticletitle{Intel Quantum Simulator: A cloud-ready
  high-performance simulator of quantum circuits}.
\newblock \bibinfo{journal}{\emph{Quantum Science \& Technology}}
  \bibinfo{volume}{5}, \bibinfo{number}{3} (\bibinfo{year}{2020}),
  \bibinfo{pages}{034007}.
\newblock


\bibitem[H{\"a}ner and Steiger(2017)]%
        {haner20175}
\bibfield{author}{\bibinfo{person}{Thomas H{\"a}ner} {and}
  \bibinfo{person}{Damian~S. Steiger}.} \bibinfo{year}{2017}\natexlab{}.
\newblock \showarticletitle{0.5 Petabyte Simulation of a 45-Qubit Quantum
  Circuit}. In \bibinfo{booktitle}{\emph{Proceedings of the International
  Conference for High Performance Computing, Networking, Storage and
  Analysis}}. \bibinfo{publisher}{ACM}, \bibinfo{address}{New York, NY, USA},
  Article \bibinfo{articleno}{33}, \bibinfo{numpages}{10}~pages.
\newblock
\urldef\tempurl%
\url{https://doi.org/10.1145/3126908.3126947}
\showDOI{\tempurl}


\bibitem[Huang et~al\mbox{.}(2026)]%
        {qfusion2026}
\bibfield{author}{\bibinfo{person}{Po-Hsuan Huang}, \bibinfo{person}{Kai-Chieh
  Chang}, \bibinfo{person}{Yu-Hui Wu}, \bibinfo{person}{Yan-Jie Wang},
  \bibinfo{person}{Yu-Tsung Wu}, \bibinfo{person}{Chia-Heng Tu}, {and}
  \bibinfo{person}{Shih-Hao Hung}.} \bibinfo{year}{2026}\natexlab{}.
\newblock \showarticletitle{{qfusion-opt}: Profile-Informed Gate Scheduling and
  Fusion Optimization for Accelerating Quantum Circuit Simulation}.
\newblock \bibinfo{journal}{\emph{IEEE Transactions on Computer-Aided Design of
  Integrated Circuits and Systems}} (\bibinfo{year}{2026}).
\newblock
\urldef\tempurl%
\url{https://doi.org/10.1109/tcad.2026.3680784}
\showDOI{\tempurl}
\newblock
\shownote{Early access}.


\bibitem[IBM(2026)]%
        {qiskit-circuit-library}
\bibfield{author}{\bibinfo{person}{IBM}.} \bibinfo{year}{2026}\natexlab{}.
\newblock \bibinfo{title}{Circuit Library}.
\newblock
\newblock
\urldef\tempurl%
\url{https://quantum.cloud.ibm.com/docs/en/api/qiskit/circuit_library}
\showURL{%
\tempurl}


\bibitem[Ittah et~al\mbox{.}(2024)]%
        {catalyst2024}
\bibfield{author}{\bibinfo{person}{David Ittah}, \bibinfo{person}{Ali Asadi},
  \bibinfo{person}{Erick~Ochoa Lopez}, \bibinfo{person}{Sergei Mironov},
  \bibinfo{person}{Samuel Banning}, \bibinfo{person}{Romain Moyard},
  \bibinfo{person}{Mai~Jacob Peng}, {and} \bibinfo{person}{Josh Izaac}.}
  \bibinfo{year}{2024}\natexlab{}.
\newblock \showarticletitle{{Catalyst}: A Python {JIT} Compiler for
  Auto-Differentiable Hybrid Quantum Programs}.
\newblock \bibinfo{journal}{\emph{Journal of Open Source Software}}
  \bibinfo{volume}{9}, \bibinfo{number}{99} (\bibinfo{year}{2024}),
  \bibinfo{pages}{6720}.
\newblock
\urldef\tempurl%
\url{https://doi.org/10.21105/joss.06720}
\showDOI{\tempurl}


\bibitem[Jiang et~al\mbox{.}(2025)]%
        {bqsim2025}
\bibfield{author}{\bibinfo{person}{Shui Jiang}, \bibinfo{person}{Yi-Hua Chung},
  \bibinfo{person}{Chih-Chun Chang}, \bibinfo{person}{Tsung-Yi Ho}, {and}
  \bibinfo{person}{Tsung-Wei Huang}.} \bibinfo{year}{2025}\natexlab{}.
\newblock \showarticletitle{{BQSim}: {GPU}-accelerated Batch Quantum Circuit
  Simulation Using Decision Diagram}. In \bibinfo{booktitle}{\emph{Proceedings
  of the 30th ACM International Conference on Architectural Support for
  Programming Languages and Operating Systems, Volume 2}}.
  \bibinfo{publisher}{ACM}, \bibinfo{address}{New York, NY, USA},
  \bibinfo{pages}{79--94}.
\newblock
\urldef\tempurl%
\url{https://doi.org/10.1145/3676641.3715984}
\showDOI{\tempurl}


\bibitem[Jones et~al\mbox{.}(2019)]%
        {jones2019quest}
\bibfield{author}{\bibinfo{person}{Tyson Jones}, \bibinfo{person}{Anna Brown},
  \bibinfo{person}{Ian Bush}, {and} \bibinfo{person}{Simon Benjamin}.}
  \bibinfo{year}{2019}\natexlab{}.
\newblock \showarticletitle{{QuEST} and High Performance Simulation of Quantum
  Computers}.
\newblock \bibinfo{journal}{\emph{Scientific Reports}}  \bibinfo{volume}{9}
  (\bibinfo{year}{2019}), \bibinfo{pages}{10736}.
\newblock
\urldef\tempurl%
\url{https://doi.org/10.1038/s41598-019-47174-9}
\showDOI{\tempurl}


\bibitem[Jones and Gacon(2020)]%
        {jones2020adjoint}
\bibfield{author}{\bibinfo{person}{Tyson Jones} {and} \bibinfo{person}{Julien
  Gacon}.} \bibinfo{year}{2020}\natexlab{}.
\newblock \bibinfo{title}{Efficient Calculation of Gradients in Classical
  Simulations of Variational Quantum Algorithms}.
\newblock
\newblock
\urldef\tempurl%
\url{https://doi.org/10.48550/arXiv.2009.02823}
\showDOI{\tempurl}
\showeprint[arxiv]{2009.02823}~[quant-ph]


\bibitem[Kirkpatrick et~al\mbox{.}(1983)]%
        {sa}
\bibfield{author}{\bibinfo{person}{S. Kirkpatrick}, \bibinfo{person}{C.~D.
  Gelatt}, {and} \bibinfo{person}{M.~P. Vecchi}.}
  \bibinfo{year}{1983}\natexlab{}.
\newblock \showarticletitle{Optimization by Simulated Annealing}.
\newblock \bibinfo{journal}{\emph{Science}} \bibinfo{volume}{220},
  \bibinfo{number}{4598} (\bibinfo{year}{1983}), \bibinfo{pages}{671--680}.
\newblock
\urldef\tempurl%
\url{https://doi.org/10.1126/science.220.4598.671}
\showDOI{\tempurl}
\showeprint{https://www.science.org/doi/pdf/10.1126/science.220.4598.671}


\bibitem[Kuzmin et~al\mbox{.}(2025)]%
        {tqml2025}
\bibfield{author}{\bibinfo{person}{Viacheslav Kuzmin}, \bibinfo{person}{Basil
  Kyriacou}, \bibinfo{person}{Tatjana Protasevich}, \bibinfo{person}{Mateusz
  Papierz}, \bibinfo{person}{Mo Kordzanganeh}, {and} \bibinfo{person}{Alexey
  Melnikov}.} \bibinfo{year}{2025}\natexlab{}.
\newblock \bibinfo{title}{{TQml Simulator}: Optimized Simulation of Quantum
  Machine Learning}.
\newblock
\newblock
\urldef\tempurl%
\url{https://doi.org/10.48550/arXiv.2506.04891}
\showDOI{\tempurl}
\showeprint[arxiv]{2506.04891}~[quant-ph]


\bibitem[Kwon et~al\mbox{.}(2020)]%
        {kwon2020nimble}
\bibfield{author}{\bibinfo{person}{Woosuk Kwon}, \bibinfo{person}{Gyeong-In
  Yu}, \bibinfo{person}{Eunji Jeong}, {and} \bibinfo{person}{Byung-Gon Chun}.}
  \bibinfo{year}{2020}\natexlab{}.
\newblock \showarticletitle{Nimble: Lightweight and parallel gpu task
  scheduling for deep learning}.
\newblock \bibinfo{journal}{\emph{Advances in Neural Information Processing
  Systems}}  \bibinfo{volume}{33} (\bibinfo{year}{2020}),
  \bibinfo{pages}{8343--8354}.
\newblock


\bibitem[Lee(2026)]%
        {pennylane-support-adjoint}
\bibfield{author}{\bibinfo{person}{Christina Lee}.}
  \bibinfo{year}{2026}\natexlab{}.
\newblock \bibinfo{title}{Adjoint Differentiation}.
\newblock
\newblock
\urldef\tempurl%
\url{https://pennylane.ai/demos/tutorial_adjoint_diff/}
\showURL{%
\tempurl}


\bibitem[Li et~al\mbox{.}(2021)]%
        {svsim2021}
\bibfield{author}{\bibinfo{person}{Ang Li}, \bibinfo{person}{Bo Fang},
  \bibinfo{person}{Christopher Granade}, \bibinfo{person}{Guen Prawiroatmodjo},
  \bibinfo{person}{Bettina Heim}, \bibinfo{person}{Martin Roetteler}, {and}
  \bibinfo{person}{Sriram Krishnamoorthy}.} \bibinfo{year}{2021}\natexlab{}.
\newblock \showarticletitle{{SV-Sim}: Scalable {PGAS}-based State Vector
  Simulation of Quantum Circuits}. In \bibinfo{booktitle}{\emph{Proceedings of
  the International Conference for High Performance Computing, Networking,
  Storage and Analysis}}. \bibinfo{publisher}{ACM}, \bibinfo{address}{New York,
  NY, USA}, Article \bibinfo{articleno}{97}, \bibinfo{numpages}{14}~pages.
\newblock
\urldef\tempurl%
\url{https://doi.org/10.1145/3458817.3476169}
\showDOI{\tempurl}


\bibitem[Li et~al\mbox{.}(2026)]%
        {dagfusion2026}
\bibfield{author}{\bibinfo{person}{Shangshu Li}, \bibinfo{person}{Zheng-An
  Wang}, {and} \bibinfo{person}{Heng Fan}.} \bibinfo{year}{2026}\natexlab{}.
\newblock \showarticletitle{{DAG}-Aware Gate Fusion for Efficient State-Vector
  Quantum-Circuit Simulation}.
\newblock \bibinfo{journal}{\emph{IEEE Transactions on Quantum Engineering}}
  \bibinfo{volume}{7} (\bibinfo{year}{2026}), \bibinfo{pages}{1--10}.
\newblock
\urldef\tempurl%
\url{https://doi.org/10.1109/tqe.2026.3705783}
\showDOI{\tempurl}


\bibitem[Liao et~al\mbox{.}(2025)]%
        {julivqc2024}
\bibfield{author}{\bibinfo{person}{Wei-You Liao}, \bibinfo{person}{Xiang Wang},
  \bibinfo{person}{Xiao-Yue Xu}, \bibinfo{person}{Chen Ding},
  \bibinfo{person}{Shuo Zhang}, \bibinfo{person}{He-Liang Huang}, {and}
  \bibinfo{person}{Chu Guo}.} \bibinfo{year}{2025}\natexlab{}.
\newblock \showarticletitle{{JuliVQC.jl}: An Efficient Variational Quantum
  Circuit Simulator for Near-Term Quantum Algorithms}.
\newblock \bibinfo{journal}{\emph{The European Physical Journal Special
  Topics}}  \bibinfo{volume}{234} (\bibinfo{year}{2025}),
  \bibinfo{pages}{6223--6239}.
\newblock
\urldef\tempurl%
\url{https://doi.org/10.1140/epjs/s11734-025-01628-6}
\showDOI{\tempurl}


\bibitem[Lu et~al\mbox{.}(2025)]%
        {cast2025}
\bibfield{author}{\bibinfo{person}{Yuncheng Lu}, \bibinfo{person}{Shuang
  Liang}, \bibinfo{person}{Hongxiang Fan}, \bibinfo{person}{Ce Guo},
  \bibinfo{person}{Wayne Luk}, {and} \bibinfo{person}{Paul H.~J. Kelly}.}
  \bibinfo{year}{2025}\natexlab{}.
\newblock \showarticletitle{Versatile Cross-platform Compilation Toolchain for
  Schr{\"o}dinger-style Quantum Circuit Simulation}. In
  \bibinfo{booktitle}{\emph{ACM/IEEE Design Automation Conference}}.
  \bibinfo{publisher}{IEEE}, \bibinfo{address}{Piscataway, NJ, USA},
  \bibinfo{pages}{1--7}.
\newblock
\urldef\tempurl%
\url{https://doi.org/10.1109/DAC63849.2025.11132539}
\showDOI{\tempurl}


\bibitem[Luo et~al\mbox{.}(2020)]%
        {luo2020yao}
\bibfield{author}{\bibinfo{person}{Xiu-Zhe Luo}, \bibinfo{person}{Jin-Guo Liu},
  \bibinfo{person}{Pan Zhang}, {and} \bibinfo{person}{Lei Wang}.}
  \bibinfo{year}{2020}\natexlab{}.
\newblock \showarticletitle{{Yao.jl}: Extensible, Efficient Framework for
  Quantum Algorithm Design}.
\newblock \bibinfo{journal}{\emph{Quantum}}  \bibinfo{volume}{4}
  (\bibinfo{year}{2020}), \bibinfo{pages}{341}.
\newblock
\urldef\tempurl%
\url{https://doi.org/10.22331/q-2020-10-11-341}
\showDOI{\tempurl}


\bibitem[Lykov et~al\mbox{.}(2021)]%
        {lykov2021performance}
\bibfield{author}{\bibinfo{person}{Danylo Lykov}, \bibinfo{person}{Angela
  Chen}, \bibinfo{person}{Huaxuan Chen}, \bibinfo{person}{Kristopher Keipert},
  \bibinfo{person}{Zheng Zhang}, \bibinfo{person}{Tom Gibbs}, {and}
  \bibinfo{person}{Yuri Alexeev}.} \bibinfo{year}{2021}\natexlab{}.
\newblock \showarticletitle{Performance evaluation and acceleration of the
  QTensor quantum circuit simulator on GPUs}. In \bibinfo{booktitle}{\emph{2021
  IEEE/ACM Second International Workshop on Quantum Computing Software (QCS)}}.
  IEEE, \bibinfo{pages}{27--34}.
\newblock


\bibitem[Lykov et~al\mbox{.}(2023)]%
        {qokit2023}
\bibfield{author}{\bibinfo{person}{Danylo Lykov}, \bibinfo{person}{Ruslan
  Shaydulin}, \bibinfo{person}{Yue Sun}, \bibinfo{person}{Yuri Alexeev}, {and}
  \bibinfo{person}{Marco Pistoia}.} \bibinfo{year}{2023}\natexlab{}.
\newblock \showarticletitle{Fast Simulation of High-Depth {QAOA} Circuits}. In
  \bibinfo{booktitle}{\emph{IEEE/ACM International Workshop on Quantum
  Computing Software}}. \bibinfo{publisher}{ACM}, \bibinfo{address}{New York,
  NY, USA}, \bibinfo{pages}{1443--1451}.
\newblock
\urldef\tempurl%
\url{https://doi.org/10.1145/3624062.3624216}
\showDOI{\tempurl}


\bibitem[Meyer et~al\mbox{.}(2023a)]%
        {PRXQuantum.4.010328}
\bibfield{author}{\bibinfo{person}{Johannes~Jakob Meyer},
  \bibinfo{person}{Marian Mularski}, \bibinfo{person}{Elies Gil-Fuster},
  \bibinfo{person}{Antonio~Anna Mele}, \bibinfo{person}{Francesco Arzani},
  \bibinfo{person}{Alissa Wilms}, {and} \bibinfo{person}{Jens Eisert}.}
  \bibinfo{year}{2023}\natexlab{a}.
\newblock \showarticletitle{Exploiting Symmetry in Variational Quantum Machine
  Learning}.
\newblock \bibinfo{journal}{\emph{PRX Quantum}}  \bibinfo{volume}{4}
  (\bibinfo{date}{Mar} \bibinfo{year}{2023}), \bibinfo{pages}{010328}.
\newblock
Issue 1.


\bibitem[Meyer et~al\mbox{.}(2023b)]%
        {meyer2023quantum}
\bibfield{author}{\bibinfo{person}{Nico Meyer}, \bibinfo{person}{Daniel
  Scherer}, \bibinfo{person}{Axel Plinge}, \bibinfo{person}{Christopher
  Mutschler}, {and} \bibinfo{person}{Michael Hartmann}.}
  \bibinfo{year}{2023}\natexlab{b}.
\newblock \showarticletitle{Quantum policy gradient algorithm with optimized
  action decoding}. In \bibinfo{booktitle}{\emph{International Conference on
  Machine Learning}}. PMLR, \bibinfo{pages}{24592--24613}.
\newblock


\bibitem[Moses et~al\mbox{.}(2021)]%
        {enzyme2021}
\bibfield{author}{\bibinfo{person}{William~S. Moses}, \bibinfo{person}{Valentin
  Churavy}, \bibinfo{person}{Ludger Paehler}, \bibinfo{person}{Jan
  H{\"u}ckelheim}, \bibinfo{person}{Sri Hari~Krishna Narayanan},
  \bibinfo{person}{Michel Schanen}, {and} \bibinfo{person}{Johannes Doerfert}.}
  \bibinfo{year}{2021}\natexlab{}.
\newblock \showarticletitle{Reverse-Mode Automatic Differentiation and
  Optimization of {GPU} Kernels via {Enzyme}}. In
  \bibinfo{booktitle}{\emph{Proceedings of the International Conference for
  High Performance Computing, Networking, Storage and Analysis}}.
  \bibinfo{publisher}{ACM}, \bibinfo{address}{New York, NY, USA}, Article
  \bibinfo{articleno}{61}, \bibinfo{numpages}{16}~pages.
\newblock
\urldef\tempurl%
\url{https://doi.org/10.1145/3458817.3476165}
\showDOI{\tempurl}


\bibitem[NVIDIA(2026a)]%
        {ncu}
\bibfield{author}{\bibinfo{person}{NVIDIA}.} \bibinfo{year}{2026}\natexlab{a}.
\newblock \bibinfo{title}{NVIDIA Nsight Compute}.
\newblock
\newblock
\urldef\tempurl%
\url{https://developer.nvidia.com/nsight-compute}
\showURL{%
\tempurl}


\bibitem[NVIDIA(2026b)]%
        {nsys}
\bibfield{author}{\bibinfo{person}{NVIDIA}.} \bibinfo{year}{2026}\natexlab{b}.
\newblock \bibinfo{title}{NVIDIA Nsight Systems}.
\newblock
\newblock
\urldef\tempurl%
\url{https://developer.nvidia.com/nsight-systems}
\showURL{%
\tempurl}


\bibitem[{PennyLane Developers}(2026)]%
        {pennylane-lightning-docs}
\bibfield{author}{\bibinfo{person}{{PennyLane Developers}}.}
  \bibinfo{year}{2026}\natexlab{}.
\newblock \bibinfo{title}{Lightning plugins}.
\newblock
\newblock
\urldef\tempurl%
\url{https://docs.pennylane.ai/projects/lightning/}
\showURL{%
\tempurl}


\bibitem[P{\'e}rez-Salinas et~al\mbox{.}(2020)]%
        {perez2020data}
\bibfield{author}{\bibinfo{person}{Adri{\'a}n P{\'e}rez-Salinas},
  \bibinfo{person}{Alba Cervera-Lierta}, \bibinfo{person}{Elies Gil-Fuster},
  {and} \bibinfo{person}{Jos{\'e}~I Latorre}.} \bibinfo{year}{2020}\natexlab{}.
\newblock \showarticletitle{Data re-uploading for a universal quantum
  classifier}.
\newblock \bibinfo{journal}{\emph{Quantum}}  \bibinfo{volume}{4}
  (\bibinfo{year}{2020}), \bibinfo{pages}{226}.
\newblock


\bibitem[Peruzzo et~al\mbox{.}(2014)]%
        {Peruzzo2014A}
\bibfield{author}{\bibinfo{person}{Alberto Peruzzo}, \bibinfo{person}{Jarrod
  McClean}, \bibinfo{person}{Peter Shadbolt}, \bibinfo{person}{Man-Hong Yung},
  \bibinfo{person}{Xiao-Qi Zhou}, \bibinfo{person}{Peter~J. Love},
  \bibinfo{person}{Alán Aspuru-Guzik}, {and} \bibinfo{person}{Jeremy~L.
  O’Brien}.} \bibinfo{year}{2014}\natexlab{}.
\newblock \showarticletitle{A variational eigenvalue solver on a photonic
  quantum processor}.
\newblock \bibinfo{journal}{\emph{Nature Communications}} \bibinfo{volume}{5},
  \bibinfo{number}{1} (\bibinfo{year}{2014}), \bibinfo{pages}{4213}.
\newblock
\showISBNx{2041-1723}
\urldef\tempurl%
\url{https://doi.org/10.1038/ncomms5213}
\showDOI{\tempurl}


\bibitem[Raedt et~al\mbox{.}(2007)]%
        {deraedt2007simulator}
\bibfield{author}{\bibinfo{person}{Koen~De Raedt}, \bibinfo{person}{Kristel
  Michielsen}, \bibinfo{person}{Hans~De Raedt}, \bibinfo{person}{Binh Trieu},
  \bibinfo{person}{Guido Arnold}, \bibinfo{person}{Marcus Richter},
  \bibinfo{person}{Thomas Lippert}, \bibinfo{person}{Hiroshi Watanabe}, {and}
  \bibinfo{person}{Nobuyasu Ito}.} \bibinfo{year}{2007}\natexlab{}.
\newblock \showarticletitle{Massively Parallel Quantum Computer Simulator}.
\newblock \bibinfo{journal}{\emph{Computer Physics Communications}}
  \bibinfo{volume}{176}, \bibinfo{number}{2} (\bibinfo{year}{2007}),
  \bibinfo{pages}{121--136}.
\newblock
\urldef\tempurl%
\url{https://doi.org/10.1016/j.cpc.2006.08.007}
\showDOI{\tempurl}


\bibitem[Rumelhart et~al\mbox{.}(1986)]%
        {Rumelhart1986Learning}
\bibfield{author}{\bibinfo{person}{David~E. Rumelhart},
  \bibinfo{person}{Geoffrey~E. Hinton}, {and} \bibinfo{person}{Ronald~J.
  Williams}.} \bibinfo{year}{1986}\natexlab{}.
\newblock \showarticletitle{Learning representations by back-propagating
  errors}.
\newblock \bibinfo{journal}{\emph{Nature}} \bibinfo{volume}{323},
  \bibinfo{number}{6088} (\bibinfo{year}{1986}), \bibinfo{pages}{533--536}.
\newblock
\showISBNx{1476-4687}
\urldef\tempurl%
\url{https://doi.org/10.1038/323533a0}
\showDOI{\tempurl}


\bibitem[Schatzki et~al\mbox{.}(2024)]%
        {Schatzki2024}
\bibfield{author}{\bibinfo{person}{Louis Schatzki}, \bibinfo{person}{Mart{\'i}n
  Larocca}, \bibinfo{person}{Quynh~T. Nguyen},
  \bibinfo{person}{Fr{\'e}d{\'e}ric Sauvage}, {and} \bibinfo{person}{M.
  Cerezo}.} \bibinfo{year}{2024}\natexlab{}.
\newblock \showarticletitle{Theoretical guarantees for permutation-equivariant
  quantum neural networks}.
\newblock \bibinfo{journal}{\emph{npj Quantum Information}}
  \bibinfo{volume}{10}, \bibinfo{number}{1} (\bibinfo{date}{22 Jan}
  \bibinfo{year}{2024}), \bibinfo{pages}{12}.
\newblock
\showISSN{2056-6387}


\bibitem[Schuld et~al\mbox{.}(2019)]%
        {schuld2019parameter}
\bibfield{author}{\bibinfo{person}{Maria Schuld}, \bibinfo{person}{Ville
  Bergholm}, \bibinfo{person}{Christian Gogolin}, \bibinfo{person}{Josh Izaac},
  {and} \bibinfo{person}{Nathan Killoran}.} \bibinfo{year}{2019}\natexlab{}.
\newblock \showarticletitle{Evaluating Analytic Gradients on Quantum Hardware}.
\newblock \bibinfo{journal}{\emph{Physical Review A}} \bibinfo{volume}{99},
  \bibinfo{number}{3} (\bibinfo{year}{2019}), \bibinfo{pages}{032331}.
\newblock
\urldef\tempurl%
\url{https://doi.org/10.1103/PhysRevA.99.032331}
\showDOI{\tempurl}


\bibitem[Stein et~al\mbox{.}(2024)]%
        {stein2024cuaoa}
\bibfield{author}{\bibinfo{person}{Jonas Stein}, \bibinfo{person}{Jonas
  Blenninger}, \bibinfo{person}{David Bucher}, \bibinfo{person}{Peter~J Eder},
  \bibinfo{person}{Elif {\c{C}}etiner}, \bibinfo{person}{Maximilian Zorn},
  {and} \bibinfo{person}{Claudia Linnhoff-Popien}.}
  \bibinfo{year}{2024}\natexlab{}.
\newblock \showarticletitle{Cuaoa: A novel cuda-accelerated simulation
  framework for the qaoa}. In \bibinfo{booktitle}{\emph{2024 IEEE International
  Conference on Quantum Computing and Engineering (QCE)}},
  Vol.~\bibinfo{volume}{2}. IEEE, \bibinfo{pages}{280--285}.
\newblock
\urldef\tempurl%
\url{https://doi.org/10.1109/QCE60285.2024.10292}
\showDOI{\tempurl}


\bibitem[Stenzel et~al\mbox{.}(2025)]%
        {qandle2025}
\bibfield{author}{\bibinfo{person}{Gerhard Stenzel}, \bibinfo{person}{Sebastian
  Zielinski}, \bibinfo{person}{Michael K{\"o}lle}, \bibinfo{person}{Philipp
  Altmann}, \bibinfo{person}{Jonas N{\"u}{\ss}lein}, {and}
  \bibinfo{person}{Thomas Gabor}.} \bibinfo{year}{2025}\natexlab{}.
\newblock \showarticletitle{{Qandle}: Accelerating State Vector Simulation
  Using Gate-Matrix Caching and Circuit Splitting}. In
  \bibinfo{booktitle}{\emph{Proceedings of the 17th International Conference on
  Agents and Artificial Intelligence, Volume 1: QAIO}}.
  \bibinfo{publisher}{SCITEPRESS - Science and Technology Publications},
  \bibinfo{address}{New York, NY, USA}, \bibinfo{pages}{715--723}.
\newblock
\urldef\tempurl%
\url{https://doi.org/10.5220/0013343500003890}
\showDOI{\tempurl}


\bibitem[Suzuki et~al\mbox{.}(2021)]%
        {suzuki2021qulacs}
\bibfield{author}{\bibinfo{person}{Yasunari Suzuki}, \bibinfo{person}{Yoshiaki
  Kawase}, \bibinfo{person}{Yuya Masumura}, \bibinfo{person}{Yuria Hiraga},
  \bibinfo{person}{Masahiro Nakadai}, \bibinfo{person}{Jiabao Chen},
  \bibinfo{person}{Ken~M. Nakanishi}, \bibinfo{person}{Kosuke Mitarai},
  \bibinfo{person}{Ryosuke Imai}, \bibinfo{person}{Shiro Tamiya},
  \bibinfo{person}{Takahiro Yamamoto}, \bibinfo{person}{Tennin Yan},
  \bibinfo{person}{Toru Kawakubo}, \bibinfo{person}{Yuya~O. Nakagawa},
  \bibinfo{person}{Yohei Ibe}, \bibinfo{person}{Youyuan Zhang},
  \bibinfo{person}{Hirotsugu Yamashita}, \bibinfo{person}{Hikaru Yoshimura},
  \bibinfo{person}{Akihiro Hayashi}, {and} \bibinfo{person}{Keisuke Fujii}.}
  \bibinfo{year}{2021}\natexlab{}.
\newblock \showarticletitle{{Qulacs}: A Fast and Versatile Quantum Circuit
  Simulator for Research Purpose}.
\newblock \bibinfo{journal}{\emph{Quantum}}  \bibinfo{volume}{5}
  (\bibinfo{year}{2021}), \bibinfo{pages}{559}.
\newblock
\urldef\tempurl%
\url{https://doi.org/10.22331/q-2021-10-06-559}
\showDOI{\tempurl}


\bibitem[Tan and Le(2019)]%
        {tan2019efficientnet}
\bibfield{author}{\bibinfo{person}{Mingxing Tan} {and} \bibinfo{person}{Quoc
  Le}.} \bibinfo{year}{2019}\natexlab{}.
\newblock \showarticletitle{Efficientnet: Rethinking model scaling for
  convolutional neural networks}. In \bibinfo{booktitle}{\emph{International
  conference on machine learning}}. PmLR, \bibinfo{pages}{6105--6114}.
\newblock


\bibitem[Team(2025)]%
        {qiskit-aer}
\bibfield{author}{\bibinfo{person}{Qiskit~Development Team}.}
  \bibinfo{year}{2025}\natexlab{}.
\newblock \bibinfo{title}{Qiskit Aer documentation}.
\newblock
\newblock
\urldef\tempurl%
\url{https://qiskit.github.io/qiskit-aer/}
\showURL{%
\tempurl}


\bibitem[Venev et~al\mbox{.}(2025)]%
        {qblaze2025}
\bibfield{author}{\bibinfo{person}{Hristo Venev}, \bibinfo{person}{Thien
  Udomsrirungruang}, \bibinfo{person}{Dimitar Dimitrov}, \bibinfo{person}{Timon
  Gehr}, {and} \bibinfo{person}{Martin Vechev}.}
  \bibinfo{year}{2025}\natexlab{}.
\newblock \showarticletitle{{qblaze}: An Efficient and Scalable Sparse Quantum
  Simulator}.
\newblock \bibinfo{journal}{\emph{Proceedings of the ACM on Programming
  Languages}} \bibinfo{volume}{9}, \bibinfo{number}{OOPSLA2}
  (\bibinfo{year}{2025}), \bibinfo{pages}{444--470}.
\newblock
\urldef\tempurl%
\url{https://doi.org/10.1145/3763066}
\showDOI{\tempurl}


\bibitem[Wang et~al\mbox{.}(2022)]%
        {wang2022qoc}
\bibfield{author}{\bibinfo{person}{Hanrui Wang}, \bibinfo{person}{Zirui Li},
  \bibinfo{person}{Jiaqi Gu}, \bibinfo{person}{Yongshan Ding},
  \bibinfo{person}{David~Z Pan}, {and} \bibinfo{person}{Song Han}.}
  \bibinfo{year}{2022}\natexlab{}.
\newblock \showarticletitle{Qoc: quantum on-chip training with parameter shift
  and gradient pruning}. In \bibinfo{booktitle}{\emph{Proceedings of the 59th
  ACM/IEEE design automation conference}}. \bibinfo{pages}{655--660}.
\newblock


\bibitem[Wiersema et~al\mbox{.}(2020)]%
        {PRXQuantum.1.020319}
\bibfield{author}{\bibinfo{person}{Roeland Wiersema}, \bibinfo{person}{Cunlu
  Zhou}, \bibinfo{person}{Yvette de Sereville}, \bibinfo{person}{Juan~Felipe
  Carrasquilla}, \bibinfo{person}{Yong~Baek Kim}, {and} \bibinfo{person}{Henry
  Yuen}.} \bibinfo{year}{2020}\natexlab{}.
\newblock \showarticletitle{Exploring Entanglement and Optimization within the
  Hamiltonian Variational Ansatz}.
\newblock \bibinfo{journal}{\emph{PRX Quantum}}  \bibinfo{volume}{1}
  (\bibinfo{date}{Dec} \bibinfo{year}{2020}), \bibinfo{pages}{020319}.
\newblock
Issue 2.
\urldef\tempurl%
\url{https://doi.org/10.1103/PRXQuantum.1.020319}
\showDOI{\tempurl}


\bibitem[Xu et~al\mbox{.}(2024)]%
        {atlas2024}
\bibfield{author}{\bibinfo{person}{Mingkuan Xu}, \bibinfo{person}{Shiyi Cao},
  \bibinfo{person}{Xupeng Miao}, \bibinfo{person}{Umut~A. Acar}, {and}
  \bibinfo{person}{Zhihao Jia}.} \bibinfo{year}{2024}\natexlab{}.
\newblock \showarticletitle{{Atlas}: Hierarchical Partitioning for Quantum
  Circuit Simulation on {GPUs}}. In \bibinfo{booktitle}{\emph{International
  Conference for High Performance Computing, Networking, Storage, and
  Analysis}}. \bibinfo{publisher}{IEEE}, \bibinfo{address}{Piscataway, NJ,
  USA}, Article \bibinfo{articleno}{81}, \bibinfo{numpages}{17}~pages.
\newblock
\urldef\tempurl%
\url{https://doi.org/10.1109/SC41406.2024.00087}
\showDOI{\tempurl}


\bibitem[Zhang et~al\mbox{.}(2025)]%
        {zhang2025bmqsim}
\bibfield{author}{\bibinfo{person}{Boyuan Zhang}, \bibinfo{person}{Bo Fang},
  \bibinfo{person}{Fanjiang Ye}, \bibinfo{person}{Luanzheng Guo},
  \bibinfo{person}{Fengguang Song}, \bibinfo{person}{Nathan Tallent}, {and}
  \bibinfo{person}{Dingwen Tao}.} \bibinfo{year}{2025}\natexlab{}.
\newblock \showarticletitle{BMQSim: Overcoming Memory Constraints in Quantum
  Circuit Simulation with a High-Fidelity Compression Framework}. In
  \bibinfo{booktitle}{\emph{Proceedings of the 39th ACM International
  Conference on Supercomputing}}. \bibinfo{pages}{689--704}.
\newblock


\bibitem[Zhang et~al\mbox{.}(2021)]%
        {hyquas2021}
\bibfield{author}{\bibinfo{person}{Chen Zhang}, \bibinfo{person}{Zeyu Song},
  \bibinfo{person}{Haojie Wang}, \bibinfo{person}{Kaiyuan Rong}, {and}
  \bibinfo{person}{Jidong Zhai}.} \bibinfo{year}{2021}\natexlab{}.
\newblock \showarticletitle{{HyQuas}: Hybrid Partitioner Based Quantum Circuit
  Simulation System on {GPU}}. In \bibinfo{booktitle}{\emph{International
  Conference on Supercomputing}}. \bibinfo{publisher}{ACM},
  \bibinfo{address}{New York, NY, USA}, \bibinfo{pages}{443--454}.
\newblock
\urldef\tempurl%
\url{https://doi.org/10.1145/3447818.3460357}
\showDOI{\tempurl}


\bibitem[Zhang et~al\mbox{.}(2022)]%
        {uniq2022}
\bibfield{author}{\bibinfo{person}{Chen Zhang}, \bibinfo{person}{Haojie Wang},
  \bibinfo{person}{Zixuan Ma}, \bibinfo{person}{Lei Xie}, \bibinfo{person}{Zeyu
  Song}, {and} \bibinfo{person}{Jidong Zhai}.} \bibinfo{year}{2022}\natexlab{}.
\newblock \showarticletitle{{UniQ}: A Unified Programming Model for Efficient
  Quantum Circuit Simulation}. In \bibinfo{booktitle}{\emph{SC22: International
  Conference for High Performance Computing, Networking, Storage and
  Analysis}}. \bibinfo{publisher}{IEEE}, \bibinfo{address}{Piscataway, NJ,
  USA}, \bibinfo{pages}{1--16}.
\newblock
\urldef\tempurl%
\url{https://doi.org/10.1109/sc41404.2022.00054}
\showDOI{\tempurl}


\bibitem[Zhang et~al\mbox{.}(2026)]%
        {zhang2026aqer}
\bibfield{author}{\bibinfo{person}{Kaining Zhang}, \bibinfo{person}{Xinbiao
  Wang}, \bibinfo{person}{Yuxuan Du}, \bibinfo{person}{Min-Hsiu Hsieh}, {and}
  \bibinfo{person}{Dacheng Tao}.} \bibinfo{year}{2026}\natexlab{}.
\newblock \bibinfo{title}{{AQER}: A Scalable and Efficient Data Loader for
  Digital Quantum Computers}.
\newblock
\newblock
\urldef\tempurl%
\url{https://doi.org/10.48550/arXiv.2602.02165}
\showDOI{\tempurl}
\showeprint[arxiv]{2602.02165}~[quant-ph]


\bibitem[Zhang et~al\mbox{.}(2023)]%
        {tensorcircuit2023}
\bibfield{author}{\bibinfo{person}{Shi-Xin Zhang}, \bibinfo{person}{Jonathan
  Allcock}, \bibinfo{person}{Zhou-Quan Wan}, \bibinfo{person}{Shuo Liu},
  \bibinfo{person}{Jiace Sun}, \bibinfo{person}{Hao Yu},
  \bibinfo{person}{Xing-Han Yang}, \bibinfo{person}{Jiezhong Qiu},
  \bibinfo{person}{Zhaofeng Ye}, \bibinfo{person}{Yu-Qin Chen},
  \bibinfo{person}{Chee-Kong Lee}, \bibinfo{person}{Yi-Cong Zheng},
  \bibinfo{person}{Shao-Kai Jian}, \bibinfo{person}{Hong Yao},
  \bibinfo{person}{Chang-Yu Hsieh}, {and} \bibinfo{person}{Shengyu Zhang}.}
  \bibinfo{year}{2023}\natexlab{}.
\newblock \showarticletitle{{TensorCircuit}: A Quantum Software Framework for
  the {NISQ} Era}.
\newblock \bibinfo{journal}{\emph{Quantum}}  \bibinfo{volume}{7}
  (\bibinfo{year}{2023}), \bibinfo{pages}{912}.
\newblock
\urldef\tempurl%
\url{https://doi.org/10.22331/q-2023-02-02-912}
\showDOI{\tempurl}


\bibitem[Zhao et~al\mbox{.}(2022)]%
        {qgpu2022}
\bibfield{author}{\bibinfo{person}{Yilun Zhao}, \bibinfo{person}{Yanan Guo},
  \bibinfo{person}{Yuan Yao}, \bibinfo{person}{Amanda Dumi},
  \bibinfo{person}{Devin~M. Mulvey}, \bibinfo{person}{Shiv Upadhyay},
  \bibinfo{person}{Youtao Zhang}, \bibinfo{person}{Kenneth~D. Jordan},
  \bibinfo{person}{Jun Yang}, {and} \bibinfo{person}{Xulong Tang}.}
  \bibinfo{year}{2022}\natexlab{}.
\newblock \showarticletitle{{Q-GPU}: A Recipe of Optimizations for Quantum
  Circuit Simulation Using {GPUs}}. In \bibinfo{booktitle}{\emph{IEEE
  International Symposium on High-Performance Computer Architecture}}.
  \bibinfo{publisher}{IEEE}, \bibinfo{address}{Piscataway, NJ, USA},
  \bibinfo{pages}{726--740}.
\newblock
\urldef\tempurl%
\url{https://doi.org/10.1109/HPCA53966.2022.00059}
\showDOI{\tempurl}


\bibitem[Zheng et~al\mbox{.}(2020)]%
        {ansor2020}
\bibfield{author}{\bibinfo{person}{Lianmin Zheng}, \bibinfo{person}{Chengfan
  Jia}, \bibinfo{person}{Minmin Sun}, \bibinfo{person}{Zhao Wu},
  \bibinfo{person}{Cody~Hao Yu}, \bibinfo{person}{Ameer Haj-Ali},
  \bibinfo{person}{Yida Wang}, \bibinfo{person}{Jun Yang},
  \bibinfo{person}{Danyang Zhuo}, \bibinfo{person}{Koushik Sen},
  \bibinfo{person}{Joseph~E. Gonzalez}, {and} \bibinfo{person}{Ion Stoica}.}
  \bibinfo{year}{2020}\natexlab{}.
\newblock \showarticletitle{{Ansor}: Generating High-Performance Tensor
  Programs for Deep Learning}. In \bibinfo{booktitle}{\emph{USENIX Symposium on
  Operating Systems Design and Implementation}}. \bibinfo{publisher}{USENIX
  Association}, \bibinfo{address}{USA}, \bibinfo{pages}{863--879}.
\newblock
\urldef\tempurl%
\url{https://www.usenix.org/conference/osdi20/presentation/zheng}
\showURL{%
\tempurl}


\bibitem[Zhu et~al\mbox{.}(2022)]%
        {roller2022}
\bibfield{author}{\bibinfo{person}{Hongyu Zhu}, \bibinfo{person}{Ruofan Wu},
  \bibinfo{person}{Yijia Diao}, \bibinfo{person}{Shanbin Ke},
  \bibinfo{person}{Haoyu Li}, \bibinfo{person}{Chen Zhang},
  \bibinfo{person}{Jilong Xue}, \bibinfo{person}{Lingxiao Ma},
  \bibinfo{person}{Yuqing Xia}, \bibinfo{person}{Wei Cui}, \bibinfo{person}{Fan
  Yang}, \bibinfo{person}{Mao Yang}, \bibinfo{person}{Lidong Zhou},
  \bibinfo{person}{Asaf Cidon}, {and} \bibinfo{person}{Gennady Pekhimenko}.}
  \bibinfo{year}{2022}\natexlab{}.
\newblock \showarticletitle{{ROLLER}: Fast and Efficient Tensor Compilation for
  Deep Learning}. In \bibinfo{booktitle}{\emph{USENIX Symposium on Operating
  Systems Design and Implementation}}. \bibinfo{publisher}{USENIX Association},
  \bibinfo{address}{Carlsbad, CA, USA}, \bibinfo{pages}{233--248}.
\newblock
\urldef\tempurl%
\url{https://www.usenix.org/conference/osdi22/presentation/zhu}
\showURL{%
\tempurl}


\end{thebibliography}
